\documentclass[a4paper,11pt]{article}

\usepackage{jheppub} 
\usepackage{lineno}
\usepackage{physics}
\usepackage{bm}
\usepackage{comment}
\usepackage[all]{xy}
\usepackage{tikz-cd}

\usepackage{amsthm}
\newtheorem{theorem}{Theorem}  
\newtheorem{lemma}[theorem]{Lemma}      

\theoremstyle{definition}

\newtheorem{remark}[theorem]{Remark}

\preprint{%
\begin{flushright}
RIKEN-iTHEMS-Report-26\\
UT-Komaba/26-4
\end{flushright}
}

\title{Exact $\mathrm{SL}(2,\mathbb{Z})$-Structure of Lattice Maxwell Theory with $\theta$-term in Modified Villain Formulation}

\author[a]{Shoto Aoki}
\emailAdd{shoto.aoki@riken.jp}
\affiliation[a]{Interdisciplinary Theoretical and Mathematical Sciences Program (iTHEMS), RIKEN, Wako 351-0198, Japan}

\author[b]{Yoshio Kikukawa}
\emailAdd{kikukawa@hep1.c.u-tokyo.ac.jp}
\affiliation[b]{Graduate School of Arts and Sciences, University of Tokyo, Komaba, Meguro-ku, Tokyo 153-8902, Japan}

\author[b]{Toshinari Takemoto}
\emailAdd{takemoto@hep1.c.u-tokyo.ac.jp}

\abstract{

We study the duality of lattice Maxwell theory in the modified Villain formulation, employing an ultra-local action with a theta term. Although this action is known to become non ultra-local through the Poisson resummation formula, we show that this non ultra-locality can be removed by incorporating a non-local transformation procedure into the definition of the $\mathcal{S}$-transformation. As a result, the ultra-local action with a theta term exhibits an exact $\mathrm{SL}(2,\mathbb{Z})$-duality. We further analyze the $\mathrm{SL}(2,\mathbb{Z})$-structure of Wilson and ’t Hooft loops, demonstrating that they transform properly up to a nontrivial phase factor arising from the nontrivial self-linking of the loops. This effect originates from the non-local transformation procedure in the $\mathcal{S}$-transformation. Remarkably, the resulting $\mathrm{SL}(2,\mathbb{Z})$-structure closely resembles that of non-spin Maxwell theory.

}

\begin{document}
\maketitle
\flushbottom

\section{Introduction}
\label{sec: intro}


Four dimensional Maxwell theory is one of the fundamental theories, which describes dynamics of fields around electrically and magnetically charged objects. These objects interact with each other via the electric and magnetic fields generated by them. The physical phenomena obey four equations, known as Maxwell's equations. As the roles of the electric and magnetic fields are very similar, these equations are invariant under the interchange of these two fields. Remarkably, the coupling constant changes to the inverse. This implies that the strong coupling theory is equivalent to the weak coupling theory. This structure is called $\mathcal{S}$-duality or self-duality.


The inner product of the electric and magnetic fields is also invariant under the exchange of them. This term is well known as $\theta$-term and takes an integer topological value. By adding the $\theta$-term to the action, the theory enjoys the $ 2\pi $ periodicity of $\theta$. This structure is called $\mathcal{T}$-duality. In the presence of a monopole, an electric charge is induced from the $\theta$-term, and the monopole becomes a dyon. This is the famous Witten effect \cite{Witten:1979ey}.


$\mathcal{S}$ and $\mathcal{T}$-dualities are generators of $\mathrm{SL}(2,\mathbb{Z})$-duality. The duality has been discovered in various theories, including not only Abelian gauge theories \cite{Witten:1995gf,Lozano:1995aq, Metlitski:2015yqa} but also non-Abelian Yang-Mills theories such as Seiberg-Witten theory \cite{Seiberg:1994rs,Seiberg:1994aj} and $\mathcal{N}=4$ super Yang-Mills theory \cite{Montonen:1977sn, Aharony:2013hda}. The existence of $\mathrm{SL}(2,\mathbb{Z})$-duality provides a powerful framework for classifying the topological structures and phases of the theory, and understanding the non-perturbative effects. 
 


In recent years, attempts have been made to implement the $\mathrm{SL}(2,\mathbb{Z})$ in lattice theories \cite{Cardy:1981fd, Cardy:1981qy,Honda:2020txe, Hayashi:2022fkw}, and the modified Villain formulation \cite{ Katayama:2025pmz, Anosova:2022cjm, Anosova:2022yqx, Fazza:2022fss, Gorantla:2021svj, Yoneda:2022qpj} is expected to exhibit the duality. In the Villain formulation \cite{Villain:1974ir}, a non-compact $1$-form gauge field $A^e$ is placed on links, and $\mathbb{Z}$-valued $2$-form gauge field $n$ on plaquettes. The $2$-form field is related to the degrees of freedoms of monopole, and makes the $1$-form field compact via electric $\mathbb{Z}$ $1$-form symmetry. In the modified Villain formulation, the dynamics of the monopole is eliminated from the theory by imposing $dn=0$. Then, a integer valued topological charge $Q_0$ is admitted in a natural way \cite{Sulejmanpasic:2019ytl},
\begin{align}
    Q_0[dA^e+2\pi n]= \frac{1}{8\pi^2} \sum_x (dA^e+2\pi n)\cup (dA^e+2\pi n).
\end{align}


The $\mathcal{S}$-transformation in the modified Villain formulation can be obtained by applying the Poisson summation formula to the $2$-form field $n$ \cite{Anosova:2022cjm,Choi:2021kmx,Sulejmanpasic:2019ytl,Gorantla:2021svj}. The $\mathcal{T}$-transformation can be defined as shifting $\theta$ by $2\pi$. However, it is not straightforward to reconcile these two. This is because, unlike the kinetic term, the topological charge is defined by a product of diagonally shifted $2$-form fields $n$. As a result, the topological charge suffers from zero modes specific to lattices, which is related to the staggered symmetry in lattice Chern-Simons theories \cite{Jacobson:2023cmr, Jacobson:2024hov, Chen:2019mjw, Xu:2024hyo, Peng:2025nfa, Eliezer:1991qh, Berruto:2000dp}.
By applying the Poisson summation formula in the presence of the $\theta$-term, the diagonal translation contaminates the kinetic term and makes it a non-local term. In the previous work \cite{Anosova:2022cjm}, a local (but non-ultra-local) kinetic term\footnote{A term is called {\it ultra-local} when its support is a finite number of lattice sites, but does not extend to infinity. The term {\it local} means that the coupling range is infinite, but the pre-factors reduce exponentially with the lattice separation. The construction presented in \cite{Anosova:2022cjm} is similar to that of the overlap Dirac operator \cite{Neuberger:1997fp}, which is a local but non-ultra-local.} has the exact $\mathcal{S}$-duality with the $\theta$-term \footnote{
To the best of our knowledge, the action does not have $\mathcal{T}$-duality since the modified kinetic term depends on $\theta$. We actually implement the path integral of the local action \cite{Anosova:2022cjm} in App.~\ref{App: partition function}.
}.


In this paper, we prove that the ultra-local Maxwell theory in modified Villain formulation with the $\theta$-term,
\begin{align}
    S= \frac{\beta}{2} \sum_x (dA^e+2\pi n)^2 + i\theta Q_0[dA^e+2\pi n] + i A^m dn ,
\end{align}
exhibits exact $\mathrm{SL}(2,\mathbb{Z})$-duality. Namely, the non-local contamination mentioned above is merely a contribution that vanishes when the integral is completely performed. In fact, the partition function is expressed by theta functions with characteristics, and its $\mathrm{SL}(2,\mathbb{Z})$-duality is obvious. The reason why the ultra-local action is not invariant under the Poisson summation is the existence of the zero modes. Modifying the topological charge to remove the zero modes, we can prove the $\mathcal{S}$-duality already before implementing the path integral. Note that the modified topological charge is equivalent to $Q_0$ in the absence of monopoles.


Our method also works in the presence of electric Wilson loops and magnetic Wilson loops (or 't Hooft loop) by replacing the topological charge $Q_0$ by
\begin{align}
    Q[dA^e+2\pi n]= &\sum_x \qty[ \frac{1}{8\pi^2} (dA^e+2\pi n)\cup (dA^e+2\pi n) + \frac{1}{4\pi} (dA^e+2\pi n) \cup_1 dn ], \\
    =&\sum_x \qty[ \frac{1}{2\pi} dA^e \cup n + \frac{1}{2} (n\cup n +n\cup_1 dn) ]
\end{align}
where $\cup_1$ is a higher cup product \cite{Chen:2021ppt, Jacobson:2023cmr}\footnote{Our topological charge provides the lattice Chern-Simons theory discussed in \cite{Jacobson:2023cmr} at the boundary. However, its bulk theory does not have $\frac{1}{2\pi} dA^e \cup n$.}. The quadratic term of $n$ is the Pontryagin square, which characterizes the symmetry protected phases \cite{Kapustin:2014gua,Hsin:2020nts, Aharony:2013hda,Gaiotto:2017yup,Kapustin:2013qsa}. The magnetic Wilson loop with magnetic charge $q_m$ induces a monopole, and $\mathcal{T}$-transformation generates only one electric Wilson loop with electric charge $q_m$ slightly displaced relative to the magnetic Wilson loop. In contrast, the conventional one generates two electric Wilson loops with $q_m/2$ \cite{Sulejmanpasic:2019ytl, Anosova:2022cjm}. We define a dyonic Wilson loop to be consistent with the $\mathcal{T}$-duality. That is, the dyonic Wilson loop is a ribbon whose edges are electric and magnetic Wilson loops. 



The $\mathcal{S}$-transformation is defined by applying the Poisson summation and changing its framing. The Poisson summation swaps the coupling constant, and exchanges electric and magnetic Wilson loops. This means that the framing of the dyonic Wilson loop is flipped. By restoring the framing, which is obtatined by alternating between the Poisson summation and $\mathcal{T}$-transformation three times, the electric and magnetic charges are switched while preserving the definition of the dyonic Wilson loop. Then, the $\mathcal{S}$ and $\mathcal{T}$-transformations generate the $\mathrm{SL(2,\mathbb{Z})}$-structure on the dyonic Wilson loop.



We note that the topological charge can take a half-integer value in the presence of the magnetic Wilson loop\footnote{Due to the half-integer valued Pontryagin square, the periodicity of $\theta$ is essentially $4\pi$ rather than $2\pi$. This is consistent with the level of the lattice Chern-Simons theory being quantized as an even number rather than an integer \cite{Jacobson:2023cmr, Jacobson:2024hov, Chen:2019mjw, Xu:2024hyo, Peng:2025nfa, Eliezer:1991qh}.}. By shifting $\theta$, the non-trivial Pontyagin square yields to properties analogous to those of the non-spin Maxwell theory \cite{Ang:2019txy, Kan:2024fuu}\footnote{The non-spin Maxwell theory is a Maxwell theory on the non-spin oriented manifold. On the non-spin manifold, the $\theta$-term becomes half-integer, and the periodicity for $\theta$ changes to $4\pi$ due to its second Stiefel–Whitney class $w_2$. Here, we can assign $w_2$ as a $\mathbb{Z}_2$ valued $2$-form gauge field of electric and magnetic $1$-form symmetry. There are four ways to turn on the background field, which corresponds to four theories. The topological nature of them are different from each other: three theories are interchangeable in $\mathrm{SL}(2,\mathbb{Z})$ transformation, while the remaining one is invariant. Although the three are anomaly-free, the remaining one exhibits a pure gravitational anomaly \cite{Thorngren:2014pza, Wang:2013zja, Wang:2018qoy,Kravec:2014aza}. 
}. We discover that the dyonic Wilson loop acquires non-trivial sign factors under the $\mathcal{S}$ and $\mathcal{T}$-transformations, and three types of Wilson loop operators are interposed. 



The paper is organized as follows. In Sec.~\ref{sec: review}, we briefly review the Maxwell theory in the modified Villain formulation, and prove the existence of the $\mathcal{S}$-duality in the absence of monopoles. This proof is prepared so that it can be immediately extended to the case with the monopoles. In Sec.~\ref{sec: dyon}, we delve into the $\mathrm{SL}(2,\mathbb{Z})$-structure of the dyonic Wilson loop. Finally, we conclude this work in Sec.~\ref{sec: conclusion}.

The technical parts are collected in Appendices. In App.~\ref{App: Differential Form}, we summarize differential forms on the lattice. We also translate the general definition of a higher cup product \cite{Chen:2021ppt, Jacobson:2023cmr} of order one into the convention of lattice gauge theories, and derive its Leibniz rule. In App.~\ref{App: Poisson}, we prove the Poisson summation formula. In App.~\ref{App: continuum}, we briefly review the $\mathrm{SL}(2,\mathbb{Z})$-duality of pure Maxwell theory in four dimensional continuum spaces. In App.~\ref{App: partition function}, we implement the path integral and derive the partition function in terms of theta functions.

\section{$U(1)$ gauge theory without background particles}
\label{sec: review}

In this section, we briefly review the four-dimensional lattice Maxwell theory in the modified Villain formulation. Although the action is ultra-local, we show that the theory exhibits the $\mathrm{SL}(2,\mathbb{Z})$-duality. 

\subsection{Review of Modified Villain Formalism}

We briefly review the partition function on a four-dimensional Euclidean torus with the local theta term in the modified Villain formalism based on \cite{Anosova:2022cjm,Sulejmanpasic:2019ytl}. We put the lattice space $\Lambda$ on the torus with the lattice spacing $a=1$, and denote the lattice sites by integer valued coordinates $x=( x_1, x_2,x_3,x_4)$ with periodic boundary conditions $x_\mu \sim x_\mu +N$. The $x_\mu $ take values in $0 ,\cdots, N-1$. 

We assign the $U(1)$ gauge field $A^e_{x,\mu}$ to links\footnote{In the standard Wilson formalism \cite{Wilson:1974sk}, the $U(1)$ gauge field is assigned as a $U(1)$-valued function $U_{x, \mu}$ and couples to electric matters as $\phi^\ast_{x} U_{x, \mu} \psi_{x+ \hat{\mu}}$, where $\hat{\mu}$ represents the unit vector in the $\mu$-direction. We parameterize the $U(1)$-valued function as
\begin{align*}
    U_{x,\mu}= \exp(i A^e_{x,\mu}).
\end{align*}}, where the superscript $e$ means ``electric'' field. The field strength is defined as
\begin{align}
    F^e_{x,\mu \nu}=& (dA^e)_{x, \mu \nu} +2\pi n_{x_{\mu\nu}} \nonumber \\
    =& A^e_{x+\hat{ \mu}, \nu} -A^e_{x, \nu} - A^e_{x+\hat{ \nu}, \mu} +A^e_{x, \mu} + 2\pi n_{x_{\mu\nu}},
\end{align}
where $n_{x,{\mu\nu}}$ is a integer valued plaquette variable corresponding to the magnetic flux through the plaquette. The field strength is invariant under $\mathbb{Z}$ $1$-form transformation
\begin{align}
    A^e_{x, \mu} &\to A^e_{x, \mu} + 2\pi k_{x, \mu}, \\
    n_{x, \mu \nu} &\to n_{x, \mu \nu}- (dk)_{x, \mu \nu}, \label{eq: Z 1-form trsf}
\end{align}
where $k_{x,\mu} \in \mathbb{Z}$. Now we can adopt a gauge in which $A^e_{x, \mu} $ takes values in $[ -\pi, \pi )$, while $n_{x,\mu\nu}$ is an arbitrary integer. 

The $\theta$-term is given by
\begin{align}
    S_\theta= \theta Q_0[F^e],
\end{align}
where $Q_0$ is a topological charge \cite{Sulejmanpasic:2019ytl,Anosova:2022cjm}. On the lattice, the topological charge is defined by
\begin{align}
    Q_0[ F^e]&= \frac{1}{8 \pi^2} \sum_x F^e \cup F^e=\frac{1}{8 \pi^2} \sum_x F^e_{x,\mu\nu} \epsilon_{\mu\nu\rho \sigma} F^e_{x+ \hat{\mu} +\hat{\nu}, \rho\sigma} \nonumber\\
    &= \frac{1}{8 \pi^2} \sum_{x,y} \sum_{\substack{ \mu< \nu \\ \rho< \sigma }}
F^e_{x, \mu\nu} \epsilon_{x, \mu \nu |y, \rho \sigma}F^e_{y, \rho \sigma }= \frac{1}{8\pi^2}  F^e \cdot \epsilon F^e .
\end{align}
Here, we regard $F^e$ as the $6 N^4$-component vector and use the matrix description. The dot represents the inner product of the vectors. The matrix $\epsilon$ is defined by
\begin{align}
    \epsilon_{x, \mu \nu |y, \rho \sigma}= \epsilon_{ \mu \nu  \rho \sigma} \delta_{x+ \hat{\mu} +\hat{\nu} , y },
\end{align}
and satisfies the following properties,
\begin{align}
    (\epsilon^2)_{x, \mu \nu |y, \rho \sigma} &= \delta_{\mu \rho} \delta_{\nu \sigma} \delta_{x+ \hat{s},y}= \delta_{\mu \rho} \delta_{\nu \sigma} (T_{\hat{s}})_{x,y},~\\
    {}^t\epsilon&= \epsilon T_{\hat{s}}^{-1}, \\
    {}^t\epsilon \epsilon &=  \epsilon {}^t\epsilon=\delta_{\mu \rho} \delta_{\nu \sigma} \delta_{x ,y}=1.
\end{align}
with $\hat{s}= \hat{1}+ \hat{2}+ \hat{3} +\hat{4}$. We denote the translational operator $x$ to $x+\hat{s}$ by $T_{\hat{s}}$.

To understand the topological charge in more detail, we define the projection operators $P_d, ~P_\partial$ and $ P_0$ into $\text{Im}(d^{(1)}),~ \text{Im}(\partial^{(3)})$ and $\text{Ker}(d^{(2)}) \cap \text{Ker}(\partial^{(2)}) $, respectively. The superscript indicates the order of the differential forms in the domain. For a $\mathbb{C}$ valued $2$-form $\alpha_{x, \mu \nu}$, the projection operators act as
\begin{align}
\begin{aligned}
    (P_d \alpha)_{x, \mu \nu} &= \sum_{p \neq 0} \frac{e^{ipx}}{ \sqrt{N^4}} \sum_{\sigma} \frac{-f_\sigma^\ast }{\abs{f}^2} ( f_\mu {\alpha}^\prime_{ \nu \sigma }-f_\nu {\alpha}^\prime_{ \mu \sigma } ) (p), \\
    (P_\partial \alpha)_{x, \mu \nu} &= \sum_{p \neq 0} \frac{e^{ipx}}{ \sqrt{N^4}} \sum_{\sigma} \frac{f_\sigma^\ast }{\abs{f}^2} ( f_\mu {\alpha}^\prime_{ \nu \sigma }-f_\nu {\alpha}^\prime_{ \mu \sigma }  + f_\sigma {\alpha}^\prime_{ \mu \nu } ) (p), \\
    (P_0 \alpha)_{x, \mu \nu} &= \frac{1}{ \sqrt{N^4}}  {\alpha}^\prime_{ \mu \nu } (0) ,
\end{aligned} \label{eq: projective operators}
\end{align}
where $f_\mu(p)= e^{ip_\mu} -1$, $\abs{f}^2= \sum_ \sigma f^\ast_\sigma f_\sigma (p) $, and ${\alpha}^\prime_{ \mu \nu } (p)$ is the Fourier transformation of $\alpha $ determined by
\begin{align}
    \alpha^\prime_{\mu\nu}(p)=\sum_{x } \frac{e^{-ipx}}{ \sqrt{N^4}} \alpha_{x, \mu \nu}.
\end{align}
with lattice momenta $p_\mu \in \frac{2\pi}{N} \mathbb{Z}$ due to the periodic boundary condition. Note that $\alpha_{x, \mu \nu}$ can be expressed as
\begin{align}
    \alpha_{x, \mu \nu}= \sum_{p } \frac{e^{ipx}}{ \sqrt{N^4}} {\alpha}^\prime_{ \mu \nu } (p).
\end{align}

The projection operators satisfy 
\begin{itemize}
\item $1=P_d + P_\partial +P_0 $,
    \item $P_i^2=P_i,~{}^tP_i=P_i$ for $i=d, \partial,0$,
    \item $P_i P_j=0$ for $i\neq j$,
    \item $P_d \epsilon = \epsilon P_ \partial,~P_\partial \epsilon = \epsilon P_d,~ P_0 \epsilon = \epsilon P_0$
    \item For any integer vector $v$, $T_v P_0=P_0$. 
\end{itemize}
We can decompose the theta term into
\begin{align}
    Q_0[F^e]&= \frac{1}{8\pi^2}  F^e \cdot \frac{\epsilon+{}^t\epsilon}{2} F^e \nonumber \\
    &= \frac{1}{8\pi^2}  F^e \cdot ( P_d + P_\partial +P_0  ) \frac{\epsilon+{}^t\epsilon}{2} (P_d + P_\partial +P_0) F^e \nonumber \\
    &=\frac{1}{8\pi^2}  F^e \cdot \qty[  P_d  \frac{\epsilon + {}^t\epsilon  }{2}  P_\partial + P_\partial  \frac{\epsilon + {}^t\epsilon  }{2}  P_d+ P_0  \frac{\epsilon + {}^t\epsilon  }{2}  P_0 ]  F^e.
\end{align}
Here, we symmetrize the matrix. Since $P_0$ is a projection to the harmonic part, ${}^t \epsilon P_0= \epsilon P_0$. Note that $Q_0$ has zero modes specific to lattices, which is related to the staggered symmetry \cite{Jacobson:2023cmr, Jacobson:2024hov, Chen:2019mjw, Xu:2024hyo, Peng:2025nfa, Eliezer:1991qh, Berruto:2000dp}. The factor $\epsilon+{}^t\epsilon= \epsilon(1+ T^{-1}_{\hat{s}})$ can be expressed in momentum space as 
\begin{align}
   \epsilon(1+ e^{-i(p_1+p_2+p_3+p_4)}).
\end{align}
Then, the topological charge becomes zero at
\begin{align}
    p_1+p_2+p_3+p_4= \pi.
\end{align}

We impose the closedness constraint or no-monopole condition
\begin{align}
    (dn)_{x, \mu \nu \rho}=0 \label{eq: closedness condition},
\end{align}
which is the counterpart of $dF=0$ in continuum spaces. Then we find
\begin{align}
    P_\partial F^e& = P_\partial  2\pi n=0, \\
    {n^\prime}_{\mu\nu}(0) &= \sum_x \frac{1}{\sqrt{N^4}}n_{x,\mu \nu} \in \mathbb{Z},
\end{align}
and
\begin{align}
Q_0[F^e]&= \frac{1}{8\pi^2}  F^e \cdot \qty[ P_0  \frac{\epsilon + {}^t\epsilon  }{2}  P_0 ]  F^e= \frac{1}{2}  n \cdot \qty[ P_0  \frac{\epsilon + {}^t\epsilon  }{2}  P_0 ]  n \nonumber \\
&=n^\prime_{1 2}(0) n^\prime_{34}(0) -n^\prime_{13}(0) n^\prime_{24}(0) +n^\prime_{14}(0) n^\prime_{23}(0)  \in \mathbb{Z}. \label{eq: theta term without monopole}
\end{align}
Here, $n^\prime_{\mu \nu}(0)$ corresponds to the first Chern number in the $(\mu \nu)$ plain\footnote{Let $c_1$ be the first Chern number. The $\theta$-term is equivalent to $\frac{1}{2}c_1^2$ in the continuum limit.}. Remarkably, the $\theta$-term only depends on the harmonic part of the $2$-form $n$. This fact leads to 
\begin{align}
    Q_0[F^e+dB]=Q_0[F^e]
\end{align}
for an arbitrary $1$-form $B$ and $Q_0$ is topological.

We define the partition function as
\begin{align}
    Z[\beta, \theta]= & \int DA^e \sum_{\{n\}} \exp( -\frac{\beta}{2}  \sum_{x}\sum_{\mu<\nu}(F^e_{x,\mu\nu})^2  - i \theta Q_0[ F^e ]) \prod_{x} \prod_{ \mu< \nu < \rho } \delta(dn_{x, \mu \nu  \rho }), \nonumber \\
    =&\int DA^e \sum_{\{n\}} \exp( -\frac{\beta}{2}  (F^e)^2  - i \theta Q_0[ F^e ]) \prod_{x} \prod_{ \mu< \nu < \rho } \delta(dn_{x, \mu \nu  \rho }),
\end{align}
where
\begin{align}
    \int DA^e = \int_{-\pi}^\pi \prod_{x} \prod_{\mu=1}^4 \frac{dA^e_{x,\mu}}{2\pi} , \quad \sum_{ \{n\} }= \prod_x \prod_{\mu<\nu} \sum_{ n_{\mu \nu} \in \mathbb{Z} },
\end{align}
and the closedness constraint is imposed by Kronecker deltas $\delta( dn)=\delta_{dn,0}$. We omit the summation symbol $\sum$ when there is no risk of confusion. By introducing the compact $3$-form $A^m_{x, \mu \nu \rho} \in [-\pi, \pi)$, we can write the Kronecker deltas as
\begin{align}
     \prod_{x} \prod_{ \mu< \nu < \rho } \delta(dn_{x, \mu \nu  \rho }) = \int DA^{m} e^{-i A^m \cdot dn} = \int DA^{m} e^{i \frac{1}{2\pi}\partial A^m \cdot F^e} ,
\end{align}
Here, the integral measure is given by
\begin{align}
\int DA^{m}=    \int_{-\pi}^\pi \prod_{x} \prod_{\mu< \nu < \rho} \frac{dA^m_{x,\mu \nu \rho}}{2\pi}.
\end{align}
Corresponding to the electric gauge field $A^e$, we refer to $A^m$ as the magnetic gauge field. Then, we get a new expression of the partition function,
\begin{align}
    Z[\beta, \theta]=& \int DA^e DA^m \sum_{\{n\}} \exp( -\frac{\beta}{2} (F^e)^2  - i \theta Q_0[ F^e ] + i\frac{1}{2\pi} \partial A^m \cdot F^e   ) \label{eq: local partition function of Ae Am}.
\end{align}
Recalling the fact that the toplogial charge $Q_0$ becomes an integer under the closedness condition, $Z[\beta,\theta]$ is $2\pi $-periodic in the $\theta$ direction:
\begin{align}
    Z[\beta, \theta+ 2\pi]= Z[\beta, \theta] \label{eq: T duality of Z}.
\end{align}
If the partition function is invariant under $\theta \to \theta +2\pi$, the theory has $\mathcal{T}$-duality.

We perform the Poisson summation formula in App.~\ref{App: Poisson}, and show how the $\theta$-term breaks the self-duality. The partition function is written as
\begin{align}
    Z[\beta, \theta]
    =&\int DA^e DA^m \sum_{\{n\}} \exp( - \frac{\beta}{2} \sum_x \sum_{\substack{ \mu<\nu \\ \rho<\sigma }} F^e_{x, \mu \nu} M_{x, \mu \nu | y, \rho \sigma} F^e_{y,\rho \sigma}  + i\frac{1}{2\pi} \partial A^m \cdot F^e   ),
\end{align}
where 
\begin{align}
 M= & 1 +i \xi \frac{\epsilon + {}^t \epsilon }{2}  \nonumber \\
 = &1+ i \xi  \qty[  P_d  \frac{\epsilon + {}^t\epsilon  }{2}  P_\partial + P_\partial  \frac{\epsilon + {}^t\epsilon  }{2}  P_d+ P_0  \frac{\epsilon + {}^t\epsilon  }{2}  P_0 ] 
\end{align}
with $\xi= \frac{\theta }{4\pi^2 \beta}$. The matrix $M$ is invertible because
\begin{align}
    M^\ast M =& 1+ \xi^2\qty( \frac{\epsilon + {}^t \epsilon }{2} )^2 =1+ \xi^2 \frac{ T_{\hat{s}} +2 +T_{\hat{s}}^{-1} }{4} 
\end{align}
is positive definite. Then, the inverse matrix of $M$ is given by
\begin{align}
    M^{-1}=& (M^\ast M )^{-1} M^\ast= \qty( 1+ \xi^2 \frac{ T_{\hat{s}} +2 +T_{\hat{s}}^{-1} }{4} )^{-1}  \qty( 1 -i \xi \frac{\epsilon + {}^t \epsilon }{2} ).
\end{align}

Finally, we implement the Poisson summation formula in App.~\ref{App: Poisson} and get
\begin{align}
    Z[\beta,\theta] = &\sqrt{\frac{1}{2\pi \beta}}^{6N^4} \frac{1}{\sqrt{\det(M)}}\int DA^e DA^m \sum_{\{m\}} \nonumber \\
    & \times \exp( - \frac{1}{2} \frac{1}{4\pi^2 \beta} \sum_x \sum_{\substack{ \mu<\nu \\ \rho<\sigma }} F^m_{x, \mu \nu} M^{-1}_{x, \mu \nu | y, \rho \sigma} F^m_{y,\rho \sigma}- i\frac{1}{2\pi} F^m \cdot  dA^e   ) , 
    \label{eq: S duality for local}
\end{align}
where $F^m= \partial A^m +2\pi m$ represents the magnetic field strength. 
Compared to Eq. \eqref{eq: local partition function of Ae Am}, the roles of $A^e$ and $A^m$ have been swapped: integrating $A^e$ generates $\partial m=0$. Then, in the imaginary part, $P_d m$ becomes zero, and only the constant term remains. We get the quadratic term as
\begin{align}
    F^m \cdot M^{-1} F^m=& F^m \cdot (M^\ast M)^{-1}  \qty(1 -i \xi \frac{\epsilon + {}^t \epsilon }{2} ) F^m \nonumber \\
    =& F^m \cdot (M^\ast M)^{-1}  F^m -i \frac{ \xi}{1+ \xi^2} F^m \cdot P_0 \frac{\epsilon + {}^t \epsilon }{2} P_0 F^m \nonumber \\
    =&F^m \cdot (M^\ast M)^{-1}  F^m -i \frac{ 8 \pi^2 \xi}{1+ \xi^2} Q[F^m],
\end{align}
and the partition function as
\begin{align}
    Z[\beta,\theta] = &\sqrt{\frac{1}{2\pi \beta}}^{6N^4} \frac{1}{\sqrt{\det(M)}}\int DA^e DA^m \sum_{\{m\}} \nonumber \\
    & \times \exp( -  \frac{1}{8\pi^2 \beta} F^m \cdot (M^\ast M)^{-1}  F^m + i \frac{1}{8 \pi^2 \beta } \frac{ 8 \pi^2 \xi}{1+ \xi^2} Q_0[F^m]-  i\frac{1}{2\pi} F^m  \cdot dA^e   ) .
\end{align}
Namely, the $\theta$-term is invariant under the $\mathcal{S}$-duality transformation. Compared to the Eq. \eqref{eq: local partition function of Ae Am}, $\theta$ transforms to
\begin{align}
    \tilde{\theta}= - \frac{1}{8 \pi^2 \beta } \frac{ 8 \pi^2 \xi}{1+ \xi^2}= \frac{-\theta}{ (2\pi\beta)^2 + (\frac{ \theta }{2\pi})^2 } \label{eq: tilde theta},
\end{align}
which is consistent with the continuum result in App.~\ref{App: continuum}. 

On the other hand, the kinetic term turns into
\begin{align}
    \frac{\beta}{2} F^m \cdot F^m \to \frac{1}{8\pi^2 \beta} F^m \cdot (M^\ast M)^{-1}  F^m = \frac{\tilde \beta}{2} F^m \cdot \frac{1+\xi^2}{M^\ast M}  F^m, 
\end{align}
with
\begin{align}
    \tilde{\beta}=  \frac{1}{4\pi^2 \beta} \frac{1}{1+ \xi^2}=  \frac{\beta}{ (2\pi\beta)^2 + (\frac{ \theta }{2\pi})^2 } \label{eq: tilde beta}.
\end{align}
If $\theta=0$, the $\beta$ converts to the inverse $1/(4\pi^2 \beta)$ since $M^\ast M$ becomes the identity matrix. That is, the Poisson summation leads to the $\mathcal{S}$-transformation. However, in the case of $\theta \neq 0$, the ultra-local kinetic term changes to the non-local kinetic term,
\begin{align}
    \frac{\tilde \beta}{2} F^m \cdot \frac{1+\xi^2}{M^\ast M}  F^m \neq  \frac{\tilde \beta}{2} F^m \cdot F^m \label{eq: naive S transformation}.
\end{align}
That is, the ultra-local action does not seem to be invariant under the $\mathcal{S}$-transformation \cite{Anosova:2022cjm}.

\subsection{$\mathrm{SL}(2,\mathbb{Z})$-duality with $\theta$-term}

In this section, we point out that the ultra-local action has the $\mathcal{S}$-duality and the non-local contamination of the kinetic term \eqref{eq: naive S transformation} is just an appearance. In fact, the violation of $\mathcal{S}$-duality arises from 
\begin{align}
     P_d  \frac{\epsilon + {}^t\epsilon  }{2}  P_\partial + P_\partial  \frac{\epsilon + {}^t\epsilon  }{2}  P_d
\end{align}
in the topological charge. However, this contribution becomes zero in the absence of monopoles. This implies that the effect of the translational operator in $(M^\ast M)^{-1}$ can be completely eliminated by implementing the path integral.

We can evaluate the path integral as shown in App.~\ref{App: partition function} and write it down in terms of theta functions. Under the closedness condition, the $2$-form $n$ can be decomposed as
\begin{align}
    n= dg + \check{h},
\end{align}
where $g$ and $\check{h}$ are integer-valued $1$- and $2$-forms. In particular, $\check{h}_{x,12}$ is given by
\begin{align}
    \check{h}_{x, 12}=\begin{cases}
        \sum_{ y_1 y_2} n_{(y_1,y_2,x_3,x_4),12}=n^\prime_{12}(0) & (x_1=x_2=N-1) \\
        0 & (\text{Otherwise})
    \end{cases},
\end{align}
and the other parts are given by equivalent expressions. Absorbing $g$ into $A^e$, we find
\begin{align}
    Z[\beta,\theta]
=&  \frac{C}{\sqrt{\beta}^{3(N^4-1)}} \prod_{\mu<\nu} \sum_{{n}^\prime_{\mu\nu}(0) \in \mathbb{Z} }  \exp( - \frac{\beta}{2} 4\pi^2 \sum_{\mu<\nu}(n^\prime_{\mu\nu }(0))^2  - i \frac{1}{2}\theta\sum_{\substack{ \mu< \nu \\ \rho< \sigma }} n^\prime_{\mu\nu }(0) \epsilon_{\mu\nu\rho \sigma}n^\prime_{\rho \sigma }(0)  ) 
\end{align}
with a positive constant number $C$ determined by the volume of the lattice. Introducing the complex coupling constant
\begin{align}
    \tau=\frac{\theta}{2\pi}+ i 2\pi \beta,
\end{align}
we find
\begin{align}
Z[\beta,\theta]=& \frac{C}{\sqrt{\beta}^{3(N^4-1)}} \qty( ~\abs{ \vartheta\mqty[ 0\\0 ] (0,2\tau) }^2 +\abs{ \vartheta\mqty[ 1/2\\0 ] (0,2\tau) }^2~    )^3,
\end{align}
where $\vartheta$ is a theta function with characteristics defined by
\begin{align}
    \vartheta\mqty[ \alpha_1\\\alpha_2 ] (q,\tau)= \sum_{n\in \mathbb{Z}} \exp( i\pi (n+\alpha_1)^2 \tau + i 2\pi (n+\alpha_1)(q+ \alpha_2)  ) \label{eq: theta function} ,
\end{align}
with $\alpha_1,\alpha_2\in \mathbb{R}$ and $q \in \mathbb{C}$.

In this expression, the $\mathrm{SL}(2,\mathbb{Z})$-duality is obvious. By applying the Poisson summation formula, we get
\begin{align}
    \vartheta\mqty[ 0\\0 ] (0,2\tau)= \sqrt{\frac{-1}{i2\tau}} \sum_m \exp( i\pi \frac{-1}{2\tau} m^2)=\sqrt{\frac{-1}{i2\tau}} \vartheta\mqty[ 0\\0 ] (0,-\frac{1}{ 2\tau}),
\end{align}
and
\begin{align}
    \vartheta\mqty[ 1/2\\0 ] (0,2\tau)= \sqrt{\frac{-1}{i2\tau}} \sum_m \exp( i\pi \frac{-1}{2\tau} m^2 -i\pi m  ) =\sqrt{\frac{-1}{i2\tau}} \vartheta\mqty[ 0\\1/2 ] (0,-\frac{1}{2\tau})
\end{align}
where the argument of the square root is given by
\begin{align}
    -\frac{\pi}{4}< \text{arg}   \sqrt{\frac{-1}{i2\tau}}<\frac{\pi}{4}.
\end{align}
These lead to
\begin{align}
    &\abs{ \vartheta\mqty[ 0\\0 ] (0,2\tau) }^2 +\abs{ \vartheta\mqty[ 1/2\\0 ] (0,2\tau) }^2 \nonumber \\
    =&\frac{1}{2 \abs{\tau}}    \qty(~ \abs{ \vartheta\mqty[ 0\\0 ] (0,-\frac{1}{2\tau}) }^2 +\abs{ \vartheta\mqty[ 0\\1/2] (0,-\frac{1}{2\tau}) }^2~) \nonumber\\
    =&\frac{1}{2 \abs{\tau}} \sum_{m_1,m_2} \exp( i\pi \frac{-1}{2\tau} m_1^2 +i\pi \frac{1}{2\bar{\tau}} m_2^2 ) (1 +  (-1)^{m_1+m_2}  ),
\end{align}
where $\bar{\tau}$ is the complex conjugate of $\tau$. Due to the last factor, the summation is limited to $m_1+m_2 \in 2\mathbb{Z}$, i.e. both $m_1$ and $ m_2$ are either even or odd integers,
\begin{align}
    &\frac{1}{2 \abs{\tau}} \sum_{m_1,m_2} \exp( i\pi \frac{-1}{2\tau} m_1^2 +i\pi \frac{1}{2\bar{\tau}} m_2^2 ) (1 +  (-1)^{m_1+m_2}  ), \nonumber \\
    =&   \frac{1}{\abs{\tau}} \sum_{n_1,n_2} \qty( \exp( i\pi \frac{-1}{2\tau} (2n_1)^2 +i\pi \frac{1}{2\bar{\tau}} (2n_2)^2 ) + \exp( i\pi \frac{-1}{2\tau} (2n_1+1)^2 +i\pi \frac{1}{2\bar{\tau}} (2n_2+1)^2 ) ) \nonumber \\
    =&\frac{1}{\abs{\tau}}  \qty(~ \abs{ \vartheta\mqty[ 0\\0 ] (0,-2\frac{1}{\tau}) }^2 +\abs{ \vartheta\mqty[ 1/2\\ 0] (0,-2\frac{1}{\tau}) }^2~).
\end{align}
Since
\begin{align}
    -\frac{1}{\tau}= \frac{\tilde{\theta}}{2\pi} + i2\pi \tilde{\beta}, 
\end{align}
the partition function is invariant under $\mathcal{S}$-transformation,
\begin{align}
    Z[\beta,\theta]=& \frac{C}{\sqrt{\beta}^{3(N^4-1)}} \frac{1}{\abs{\tau}^3} \qty(~ \abs{ \vartheta\mqty[ 0\\0 ] (0,-2\frac{1}{\tau}) }^2 +\abs{ \vartheta\mqty[ 1/2\\ 0] (0,-2\frac{1}{\tau}) }^2~)^3 \nonumber  \\
    =&\sqrt{ \frac{\tilde{\beta}}{\beta} }^{3N^4} Z[\tilde{\beta},\tilde{\theta}].\label{eq: S duality of Z}
\end{align}
Here, we use $\frac{1}{\abs{\tau} }= \sqrt{ \frac{\tilde{\beta}}{\beta}  }$. Remarkably, only the harmonic part of the $2$-form $n$ plays an important role for the duality, which is consistent with continuum theories \cite{Witten:1995gf}.

In conclusion, the partition function has two invariances,
\begin{align}
    \mathcal{T}:&\tau \to \frac{\theta+2\pi}{2\pi}+ i 2\pi \beta= \tau+1,  \\
    \mathcal{S}:&\tau \to  \frac{\tilde{\theta}}{2\pi}+ i 2\pi \tilde{\beta}= - \frac{1}{\tau}. 
\end{align}
The $\mathcal{S} $- and $\mathcal{T}$-transformations generate the modular group $\mathrm{SL}(2,\mathbb{Z})$ whose elements are expressed as
\begin{align}
    M= \mqty( a& b \\ c &d ) ,~ad-bc=1
\end{align}
with integer values $a,b,c$ and $d$. It acts on $\tau$ as
\begin{align}
    \tau \to M \tau= \frac{a\tau +b}{ c\tau+d}.
\end{align}
Here, $\mathcal{S} $ and $\mathcal{T}$ correspond to
\begin{align}
    S= \mqty( 0 & -1 \\ 1 & 0 ) , \quad T= \mqty( 1 & 1 \\ 0 & 1 )
\end{align}
in $\mathrm{SL}(2,\mathbb{Z})$.

We redefine the partition function as $\bar{Z}[\tau]= \sqrt{\beta}^{3N^4}Z[\beta,\theta]$, which is invariant under two dualities,
\begin{align}
    \bar{Z}[\tau]= \bar{Z}[S \tau]= \bar{Z}[- \frac{1}{\tau}],~ \bar{Z}[\tau]= \bar{Z}[T \tau]=\bar{Z}[\tau+1].  
\end{align}
According to the above arguments, $\bar{Z}(\tau)$ has the full $SL(2, \mathbb{Z})$ invariance,
\begin{align}
    \bar{Z}[\tau]= \bar{Z}[M \tau]= \bar{Z}[\frac{a\tau +b}{ c\tau +d}].
\end{align}

This observation suggests that we can prove the self-duality of the ultra-local action before implementing the path integral. Since the non-locality comes from the zero modes of the topological charge as discussed above, we modify the topological charge by
\begin{align}
    Q_{NL}[F^e]=\frac{1}{8\pi^2}  F^e \cdot \qty[  P_d  \epsilon  P_\partial + P_\partial   {}^t\epsilon   P_d+ P_0  \frac{\epsilon + {}^t\epsilon  }{2}  P_0 ]  F^e, \nonumber \\
    =Q_0[F^e]+ \frac{1}{8\pi^2}  F^e \cdot \qty[  P_d  \frac{\epsilon -{}^t\epsilon}{2}  P_\partial - P_\partial   \frac{\epsilon -{}^t\epsilon}{2}  P_d]  F^e.
\end{align}
Although the additive term is non-local, the contribution is zero under the closedness condition. That is,
\begin{align}
    Q_{NL}=Q_0
\end{align}
in the absence of monopoles, the partition function \eqref{eq: local partition function of Ae Am} can be expressed as
\begin{align}
    Z[\beta, \theta]=&\int DA^e DA^m \sum_{\{n\}} \exp( - \frac{\beta}{2}(F^e)^2 -i\theta Q_{NL}[F^e]    + i\frac{1}{2\pi} \partial A^m \cdot F^e  ) \nonumber \\
    =&\int DA^e DA^m \sum_{\{n\}} \exp( - \frac{\beta}{2} \sum_x \sum_{\substack{ \mu< \nu \\ \rho< \sigma }} F^e_{x, \mu \nu} M_{x, \mu \nu | y, \rho \sigma} F^e_{y,\rho \sigma}  + i\frac{1}{2\pi} \partial A^m \cdot F^e  )\nonumber \\
   =& \int DA^e DA^m \sum_{\{n\}} \exp( - \frac{\beta}{2} F^e \cdot M F^e  + i\frac{1}{2\pi} \partial A^m \cdot F^e  ) \label{eq: partition function}
\end{align}
where the matrix $M$ is given by\footnote{Such a matrix is also found in the $2$D model \cite{Katayama:2025pmz}.} 
\begin{align}
 M
 = &1+ i \xi  \qty[  P_d  \epsilon P_\partial + P_\partial{}^t \epsilon  P_d+ P_0  \frac{\epsilon + {}^t\epsilon  }{2}  P_0 ]. 
\end{align}
Since $M$ satisfies 
\begin{align}
    M^\ast M= 1+ \xi^2,
\end{align}
the inverse and determinant are expressed as
\begin{align}
    M^{-1}=  \frac{ M^\ast}{1+ \xi^2},~\det(M)= \sqrt{1+\xi^2}^{6N^4} 
\end{align}
Using the Poisson summation formula, we obtain
\begin{align}
   Z[\beta, \theta]
    =&\sqrt{\frac{1}{2\pi \beta}}^{6N^4} \frac{1}{\sqrt{\det(M)}}\int DA^e DA^m \sum_{\{m\}} \nonumber \\
    & \times \exp( - \frac{1}{2} \frac{1}{4\pi^2 \beta}F^m \cdot M^{-1} F^m - i\frac{1}{2\pi} F^m \cdot  dA^e   ) \nonumber \\
   = &\sqrt{ \frac{\tilde{\beta}}{\beta}}^{3N^4} \int DA^e DA^m \sum_{\{m\}} \exp( - \frac{\tilde{\beta}}{2} F^m \cdot F^m -i \tilde{\theta} Q_{NL}[F^m]  - i\frac{1}{2\pi} d A^e \cdot F^m   ) 
\end{align}
where $F^m= \partial A^m +2\pi m$ represents the magnetic field strength.
Here, the new parameters $\tilde{\beta}$ and $\tilde{\theta}$ are given by Eqs.~\eqref{eq: tilde theta} and \eqref{eq: tilde beta}.

To complete the proof of the self-duality, we introduce a dual lattice $\tilde{\Lambda}= \{ \tilde{x}=x+ \frac{1}{2} \hat{s} \mid x \in \Lambda  \}$. The differential forms on the original lattice are identified by
\begin{align}
\begin{aligned}
    A^e_{x,\mu}&= \sum_{\nu<\rho<\sigma} \epsilon_{\mu\nu\rho\sigma} \tilde{A}^e_{F(x)-\hat{\nu} -\hat{\rho}-\hat{\sigma} , \nu\rho\sigma},~\\
    m_{x,\mu\nu}&=\sum_{\rho<\sigma} \epsilon_{\mu\nu\rho\sigma} \tilde{m}_{F(x) -\hat{\rho}-\hat{\sigma} , \rho\sigma},~\\
    A^m_{x,\mu\nu \rho}&=\sum_{\sigma} \epsilon_{\mu\nu\rho\sigma} \tilde{A}^m_{F(x) -\hat{\sigma} , \sigma},
\end{aligned} \label{eq: Hodge star}
\end{align}
where $F: \Lambda \to \tilde{\Lambda},~F(x)= x+ \frac{1}{2}\hat{s}$. The derivative operators $d$ and $\partial$ are switched in the dual lattice,
\begin{align}
    (\partial A^m)_{x, \mu \nu}=\sum_{\rho<\sigma} \epsilon_{\mu\nu\rho\sigma} d\tilde{A}^m_{F(x) -\hat{\rho}-\hat{\sigma} , \rho\sigma} ,~ (d A^e)_{x, \mu \nu}=-\sum_{\rho<\sigma} \epsilon_{\mu\nu\rho\sigma} \partial \tilde{A}^e_{F(x) -\hat{\rho}-\hat{\sigma} , \rho\sigma}.
\end{align}
This leads to
\begin{align}
    (F^m)_{x,\mu\nu}=\sum_{\rho<\sigma} \epsilon_{\mu\nu\rho\sigma} (d\tilde{A}^m +2\pi \tilde{m})  _{F(x) -\hat{\rho}-\hat{\sigma} , \rho\sigma}.
\end{align}
We also find
\begin{align}
    (P_dn)_{x,\mu \nu}= \sum_{\rho< \sigma} \epsilon_{\mu\nu\rho\sigma}  (P_\partial \tilde{n})_{F(x) -\hat{\rho}-\hat{\sigma} , \rho\sigma},~
    (P_\partial n)_{x,\mu \nu}= \sum_{\rho< \sigma} \epsilon_{\mu\nu\rho\sigma}  (P_d \tilde{n})_{F(x) -\hat{\rho}-\hat{\sigma} , \rho\sigma}.
\end{align}
Thus, the non-local $\theta$-term  can be rewritten by
\begin{align}
    Q_{NL}[F^m]=& \frac{1}{8\pi^2}  {F}^m \cdot \qty[  P_d  \epsilon    P_\partial + P_\partial{}^t \epsilon P_d+ P_0  \frac{\epsilon + {}^t\epsilon  }{2}  P_0 ]  F^m \nonumber \\
    =&\frac{1}{8\pi^2}  \tilde{F}^m \cdot \qty[  P_d {}^t \epsilon    P_\partial + P_\partial \epsilon P_d+ P_0  \frac{\epsilon + {}^t\epsilon  }{2}  P_0 ]  \tilde{F}^e \nonumber \\
    =&\tilde{Q}_{NL}[\tilde{F}^m].
\end{align}
Since the translation into the dual lattice is unitary, we get
\begin{align}
    Z[\beta, \theta]=&\sqrt{ \frac{\tilde{\beta}}{\beta}}^{3N^4} \int D\tilde{A}^e D\tilde{A}^m \sum_{\{m\}} \exp( - \frac{\tilde{\beta}}{2} ( \tilde{F}^m)^2  -i \tilde{\theta} \tilde{Q}_{NL}[ \tilde{F}^m]  + i\frac{1}{2\pi} \partial \tilde{A}^e  \cdot\tilde{F}^m   ).
\end{align}
Compared to Eq.~\eqref{eq: partition function}, the role of $A^e$ and $A^m$ is swapped: integrating $\tilde{A}^e$ generates the closedness condition $d\tilde{m}=0$, and the dual magnetic field $\tilde{A}^m$ describes the dynamics. 
Under the closedness condition for $\tilde{m}$, the topological charges take the same value,
\begin{align}
   \tilde{Q}_{NL}=Q_0.
\end{align}
Thus, we prove the self-duality.

In the free case, we clarify the appropriate procedure for constructing the $\mathcal{S}$-transformation at $\theta\neq 0$. This involves replacing $Q_0$ with $Q_{NL}$ and applying the Poisson summation formula. Although $Q_{NL}$ is a non-local term, its non-local contribution vanishes due to the closedness condition. By incorporating this non-local procedure into the $\mathcal{S}$-transformation, the action remains ultra-local.

\section{$\mathrm{SL}(2,\mathbb{Z})$-structure with Loop operators}
\label{sec: dyon}


In this section, we discuss the $\mathrm{SL}(2,\mathbb{Z})$-structure in the presence of loop operators, namely electric, magnetic, and dyonic Wilson loops. We first extend the definition of the topological charge to the case including magnetic Wilson loops. Under the $\mathcal{T}$-transformation $\theta \to \theta + 2\pi$, a magnetic Wilson loop is dressed by an unsmeared electric Wilson loop. This is the well-known Witten effect \cite{Witten:1979ey}. 

The dyonic Wilson loop is then defined so as to be consistent with the Witten effect, and as a result, it requires a framing. The $\mathcal{S}$-transformation is induced by the Poisson summation formula; however, one must fix the framing of the dyonic Wilson loop. Remarkably, the resulting $\mathrm{SL}(2,\mathbb{Z})$-structure resembles that of non-spin Maxwell theory \cite{Ang:2019txy,Kan:2024fuu}.


\subsection{Topological Charge with Monopoles}

We first attempt to extend the previous argument to the case with monopoles. Since the monopoles violate the closedness condition, we generalize the definition of the topological charge $Q$. We assume the following:
\begin{enumerate}
\item $Q$ is defined as an ultra-local term.
\item $Q$ can be expressed as
\begin{align}
Q[F^e] = Q_{NL}[F^e] + \text{(quadratic terms in $dn$)}.
\end{align}
\item Under the closedness condition, $Q=Q_0$ .
\end{enumerate}
The third condition follows from the second. We find the topological term satisfying the above conditions\footnote{
The topological charge $Q$ is defined by adding the term $\frac{1}{2\pi} dA^e \cup n$ to that presented in \cite{Jacobson:2023cmr}. The term $\frac{1}{2\pi} dA^e \cup n$ encodes the Witten effect, and it is not necessary to change $A^m$ under the electric $\mathbb{Z}$ $1$-form transformation.
},
\begin{align}
    Q[F^e]= \sum_{x} \qty(\frac{1}{8\pi^2}  F^e \cup F^e+ \frac{1}{4\pi}F^e \cup_1 dn )\nonumber \\
    =  \sum_x \qty( \frac{1}{2\pi} dA^e \cup n +\frac{1}{2}( n\cup n + n\cup_1 dn  )  ).
\end{align}
Here, the higher cup product $\cup_1$ is defined in \eqref{eq: higher cup product}. By using the projection operators $P_d,~P_\partial$ and $P_0$, the topological charge can be decomposed as
\begin{align}
    Q[F^e]=&\frac{1}{8\pi^2}  F^e \cdot \qty[  P_d  \epsilon  P_\partial + P_\partial   {}^t\epsilon   P_d+ P_0  \frac{\epsilon + {}^t\epsilon  }{2}  P_0 ]  F^e + \frac{ 1}{2} \sum_x( P_\partial n \cup_1 d P_\partial n)_x \nonumber \\
    =&Q_{NL}[F^e]+ \frac{ 1}{2} \sum_x( P_\partial n \cup_1 d P_\partial n)_x .
\end{align}
Since $P_\partial n$ is proportional to $dn$, the topological charge satisfies our assumptions.

By shifting $\theta$ by $2\pi$, the difference of the $\theta$-term is given by
\begin{align}
    i 2\pi Q[F^e]= i \sum_x\qty( A^e \cup dn + \pi (n\cup n + n\cup_1 dn)).
\end{align}
The first term is interpreted as the Witten effect, and the second term is the Pontryagin square 
\begin{align}
    \mathfrak{P}(n)=  \sum_{x}(n\cup n + n\cup_1 dn ) \label{eq: Pontryagin square},   
\end{align}
which characterizes symmetry protected topological phases \cite{Kapustin:2014gua,Hsin:2020nts, Aharony:2013hda,Gaiotto:2017yup,Kapustin:2013qsa}. Under the electric one form transformation, the square can differ by an even number,
\begin{align}
    \mathfrak{P}(n-dk)=\mathfrak{P}(n) +\mathfrak{P}(dk)-2 \sum_x n \cup dk
    = \mathfrak{P}(n)+ 2\mathbb{Z}.
\end{align}

Since the Pontryagin square can take odd values, the periodicity of $\theta$ is essentially $4\pi$ rather than $2\pi$. We introduce the discrete $\theta$-term, and restrict the range of $\theta$ to $2\pi$. Then, we consider the following action,
\begin{align}
    S(\beta, \theta, p)= \frac{\beta}{2}\sum_x (dA^e+2\pi n)^2 +i \theta Q[F^e] + i{p \pi } \mathfrak{P}(n) +i A^m\cdot  dn,
\end{align}
where $p=0,1$ is a discrete theta angle. Then, sending $\theta\to \theta+2\pi$ is equivalent to\footnote{Such a phenomenon is also discovered in $U(1)/ \mathbb{Z}_N$ and $SU(N)/ \mathbb{Z}_N$ gauge theories.}.
\begin{align}
    S(\beta, \theta+ 2\pi, p)= S(\beta, \theta, p+1)+ \sum_x iA^e\cup dn.
\end{align}

By replacing $Q_{NL}$ by $\tilde{Q}_{NL}$ in the second condition, we can also define another type of the topological charge,
\begin{align}
    \tilde{Q}[F^e]=&\sum_{x} \qty(\frac{1}{8\pi^2}  F^e \cup F^e- \frac{1}{4\pi}F^e \cup_1 dn ) \nonumber \\
    =&\sum_x \qty( \frac{1}{2\pi} n \cup dA^e +\frac{1}{2}( n\cup n - n\cup_1 dn  )).
\end{align}
The property is almost the same as $Q$, and the last term is equivalent to the previous Pontryagin square modulo two. However, the position of the Wilson loop that resulted in the Witten effect is different. In terms of the projection operators, $\tilde{Q}$ can be rewritten as
\begin{align}
    \tilde{Q}[F^e]=\tilde{Q}_{NL}[F^e]-  \frac{ 1}{2} \sum_x( P_\partial n \cup_1 d P_\partial n)_x .\label{eq: tilde Q non local expression}
\end{align}

\subsection{Loop Operators}

The electric Wilson loop operator can be interpreted as the trajectory that describes the process of a pair creation and annihilation of an electric particle. Let $q_e \in \mathbb{Z}$ be an electric charge of the particle and $\gamma$ be the trajectory. $\gamma$ consists of oriented links from $x$ to $x\pm \mu$ denoted by $l=(x,\pm \mu)$. The electric Wilson loop along the contour $\gamma$ is defined by
\begin{align}
    W_e^{q_e}(\gamma) = \prod_{l\in \gamma} \exp( iq_e A^e_l),
\end{align}
where the electric field satisfies
\begin{align}
    A^e_{x,-\mu}=- A^e_{x-\hat{\mu},\mu}.
\end{align}
for the negative oriented link $l=(x,-\mu)$. It is convenient to introduce the tangent unit vector of $\gamma$ defined by\footnote{Mathematically, $\dot{\gamma}$ is the Thom class of counter $\gamma$.}
\begin{align}
    \dot{\gamma}_{x,\mu}=   \begin{cases}
+1 & \text{if } (x,\mu)\in \gamma,\\
-1 & \text{if } (x+\hat\mu,-\mu)\in \gamma,\\
0 & \text{otherwise}.
\end{cases}
\end{align}
Then, we find an expression of the electric Wilson loop,
\begin{align}
     W_e^{q_m}(\gamma) =  \exp( iq_e \sum_{x,\mu} \dot{\gamma}_{x,\mu}  A^e_{x,\mu}).
\end{align}
Note that $\dot{\gamma}$ satisfies 
\begin{align}
    \partial \dot{\gamma}=0.
\end{align}

Similarly, we also determine the magnetic Wilson loop (or 't Hooft loop). We assume that a magnetally charged particle lives in the dual lattice $\tilde{\Lambda}$ and couples to the magnetic gauge field $A^m_{x, \mu\nu\rho}= \sum \epsilon_{\mu\nu\rho \sigma} \tilde{A}^m_{F(x)-\hat{\sigma},\sigma}$. We let $\tilde{\gamma}$ be a loop along which the magnetic particle propagates. The magnetic Wilson loop along $\tilde{\gamma}$ is given by
\begin{align}
    W_m^{q_m}(\tilde{\gamma})
    =& \exp(  iq_m \sum_{\tilde{x},\mu} \dot{\tilde{\gamma}}_{\tilde{x} ,\mu }   \tilde{A}^m_{\tilde{x} ,\mu }  ) 
\end{align}
where $q_m \in \mathbb{Z}$ is the magnetic charge and $F^{-1}(\tilde{l})=(\tilde{x}-\frac{1}{2}\hat{s} , \pm \mu) $ for $\tilde{l}= (\tilde{x}, \pm \hat{\mu})$. 

The dyonic Wilson loop is related to the dyonic particle with the electric charge $q_e$ and the magnetic charge $q_m$. Here, the electric charge is located at the lattice site $x$, while the magnetic charge is at the dual lattice site $F(x)=x+\frac{\hat{s}}{2}$. The position is determined to be consistent with the Witten effect (we will see the next section). We define the dyonic Wilson loop along a contour $\gamma$ in the lattice space by
\begin{align}
W_d^{(q_e,q_m)}(\gamma)=&W_e^{q_e}(\gamma) W^{q_m}_m(F(\gamma)) \nonumber \\
=&\exp( i \sum_{x} ( q_e \dot{\gamma}  A^e +q_m \dot{\gamma}  \tilde{A}^m   ) ) \nonumber \\
=&\exp( i \sum_{x} ( q_e \dot{\gamma} A^e -q_m \dot{\gamma} \cup A^m   ) ) \label{eq: dyonic Wilson loop}.
\end{align}
Here, we set
\begin{align}
    \dot{\tilde{\gamma}}_{F(x),\mu}=\dot{\gamma}_{x,\mu} .
\end{align}

\begin{figure}
    \centering
    \includegraphics[width=\linewidth,bb=0 0 1305 570]{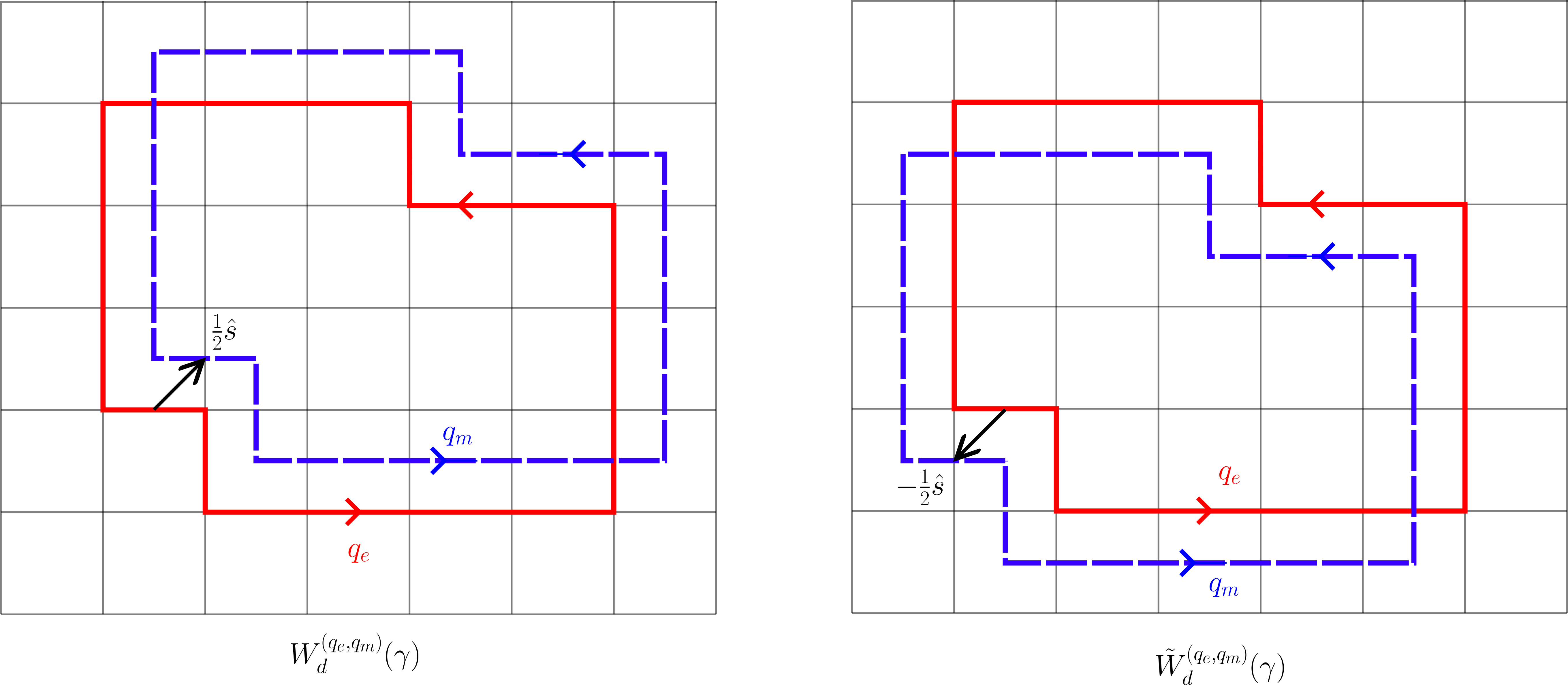}
    \caption{Left: the dyonic Wilson loop $W_d^{(q_e,q_m)}(\gamma)$ with the electric and magnetic charge $q_e$ and $q_m$, defined in Eq.~\eqref{eq: dyonic Wilson loop}. The red and blue contours represent the electric and magnetic Wilson loops, respectively. Right: the same picture of the dual dyonic Wilson loop $\tilde{W}_d^{(q_e,q_m)}({\gamma})$.
    } 
    \label{fig: dyonicWilsonloop}
\end{figure}

In the case of $\tilde{Q}$, the position of the electric and magnetic loop is interchanged. Namely, the electric charge is located at $x$ and the magnetic charge is at $\tilde{F}(x)=x-\frac{\hat{s}}{2}$. We let the dual dyonic Wilson loop be defined as
\begin{align}
\tilde{W}_d^{(q_e,q_m)}({\gamma})=&W_e^{q_e}({\gamma}) W^{q_m}_m(\tilde{F}({\gamma}))=\exp( i \sum_{x} ( q_e \dot{\gamma}  A^e +q_m  A^m \cup \dot{\gamma}  ) ) . \label{eq: dual dyonic Wilson loop}
\end{align}
We depict the two dyonic Wilson loops in Fig.~\ref{fig: dyonicWilsonloop}.

These expectation values are given by
\begin{align}
    \ev{ W_d^{(q_e,q_m)}({\gamma}) }_{\beta, \theta}^p
    = \frac{1}{Z[\beta ,\theta]} \int DA^e DA^m \sum_{\{n\}} \nonumber &\exp( -\frac{\beta}{2}  (F^e)^2  - i \theta Q[ F^e ] + i\frac{1}{2\pi} \partial A^m F^e  -ip \pi \mathfrak{P} (n) ) \\
   \times &\exp(i\sum_{x,\mu}  \qty( q_e \dot{\gamma} A^e  -  q_m \dot{\gamma} \cup A^m ) ). \label{eq: ev of dyonic loop}
\end{align}
and
\begin{align}
    \ev{ \tilde{W}_d^{(q_e,q_m)}({\gamma}) }_{\beta, \theta}^p
    = \frac{1}{Z[\beta ,\theta]} \int DA^e DA^m \sum_{\{n\}} \nonumber &\exp( -\frac{\beta}{2}  (F^e)^2  - i \theta\tilde{ Q}[ F^e ] + i\frac{1}{2\pi} \partial A^m F^e  -ip \pi \mathfrak{P} (n) ) \\
\times &\exp(i\sum_{x,\mu}  \qty( q_e \dot{\gamma} A^e  +  q_m   A^m \cup \dot{\gamma}  ) ). \label{eq: ev of dual dyonic loop}
\end{align}
Here, we replace $Q$ by $\tilde{Q}$ in the case of $\tilde{W}$.

In the rest of this section, we prove some identities of $\ev{W_d}$ and $\ev{\tilde{W}_d}$ for the fixed contour $\gamma$. To do this, we define an integer-valued $2$-form $l$ such that
\begin{align}
    \epsilon_{\mu\nu\rho \sigma} dl_{x+\hat{\mu}, \nu\rho \sigma}= -\dot\gamma_{x,\mu},
\end{align}
and we set 
\begin{align}
    u=\sum_x P_\partial l \cup_1d P_\partial l,~v= \mathfrak{P}(l).
\end{align}
Note that $u$ and $v$ are real and integer numbers, respectively. 
The construction of $l$ is as follows.
\begin{enumerate}
    \item Find an oriented surface $\Sigma$ whose boundary is $\gamma$. 
    The monopole constraint \eqref{eq: monopole constraint} ensures the existence of such a surface.
    
    \item Define a $2$-form $\tilde{l}$ such that
    \begin{align}
        \sum_{p \in \Sigma} \alpha_p = \sum_x \sum_{\mu<\nu} \alpha_{x,\mu\nu} \tilde{l}_{x,\mu\nu}
    \end{align}
    for any $2$-form $\alpha$.
    
    \item Set
    \begin{align}
        l_{x,\mu\nu} = -\sum_{\rho<\sigma} \epsilon_{\mu\nu\rho\sigma} \, \tilde{l}_{x-\hat{\rho}-\hat{\sigma},\rho\sigma}.
    \end{align}
\end{enumerate}

Then the following lemmas hold.
\begin{lemma}[$\mathcal{T}$-transformation] \label{lem: T}
    Shifting $\theta$ by $2\pi$, we have
    \begin{align}
        \ev{ {W}_d^{(q_e-q_m,q_m)}({\gamma}) }_{\beta, \theta+2\pi}^p &=\ev{ {W}_d^{(q_e,q_m)}({\gamma}) }_{\beta, \theta}^{p+1} =\ev{ {W}_d^{(q_e,q_m)}({\gamma}) }_{\beta, \theta}^{p}(-1)^{q_m^2 v},\\
        \ev{ \tilde{W}_d^{(q_e-q_m,q_m)}({\gamma}) }_{\beta, \theta+2\pi}^p &=\ev{ \tilde{W}_d^{(q_e,q_m)}({\gamma}) }_{\beta, \theta}^{p+1}=\ev{ \tilde{W}_d^{(q_e,q_m)}({\gamma}) }_{\beta, \theta}^{p}(-1)^{q_m^2 v}.
    \end{align}
    This is the Witten effect. Then $\mathcal{T}$ acts on the dyonic charge $(q_e, q_m)$ as
    \begin{align}
        \mqty( q_e\\ q_m) \mapsto \mqty( q_e-q_m \\ q_m) =\mqty(1 & -1 \\ 0 & 1) \mqty( q_e\\ q_m) \label{eq: T for charge}.
    \end{align}
\end{lemma}

\begin{lemma}[Poisson summation] \label{lem: P}
    Applying the Poisson summation formula, we get
    \begin{align}
        \ev{ {W}_d^{(q_e,q_m)}({\gamma}) }_{\beta, \theta}^p= \ev{ {\tilde{W}}_d^{(q_m,-q_e)}( F({\gamma})) }_{\tilde{\beta}, \tilde{\theta}}^p \exp( -i\theta u\frac{1}{2} q_m^2 -i\tilde{\theta} u\frac{1}{2} q_e^2 -ip\pi v(q_m^2+ q_e^2)  ) .
    \end{align}
    Here, the electric and magnetic Wilson loops are interchanged. As a result, $Q\to \tilde{Q}$ and $W_d\to \tilde{W}_d$. Since the Poisson summation changes the framing of the dyonic Wilson loop, we denote the operation of applying the Poisson summation formula by $\mathcal{P}$ and distinguish it from the $\mathcal{S}$-transformation. 

By applying the Poisson summation twice, we also find
\begin{align}
    \ev{ {W}_d^{(q_e,q_m)}({\gamma}) }_{\beta, \theta}^p=\ev{ {W}_d^{(-q_e,-q_m)}({\gamma}) }_{\beta, \theta}^p.
\end{align}
\end{lemma}

By combining the $\mathcal{T}$-transformation and $\mathcal{P}$, we can prove the following lemmas. 
\begin{lemma}[Flipping framings] \label{lem: framing}
    By flipping the framing of the dyonic Wilson loop, the Wilson loop acquires the following phase,
\begin{align}
    \ev{ \tilde{W}_d^{(q_e,q_m)}(F{\gamma}) }_{\beta, \theta}^p
    =&\ev{ {W}_d^{(q_e,q_m)}({\gamma}) }_{\beta, \theta}^p\exp( i 2\pi u(q_e+\frac{\theta}{2\pi}q_m )q_m  )\label{eq: anomaly}.
\end{align}
This process can be described by alternating $\mathcal{P}$ and $\mathcal{T}$ in three times.
\end{lemma}

\begin{lemma}[$\mathcal{S}$-transformation] \label{lem: S}
    The $\mathcal{S}$-transformation is defined by the Poisson summation and flipping the framing. Then
\begin{align}
    \ev{ {W}_d^{(q_e,q_m)}({\gamma}) }_{\beta, \theta}^p
    = &\exp(-iu\pi \frac{q^2}{\tau} - 2\pi^2 u ( \beta q_m^2 -\tilde{\beta} q_e^2) )  (-1)^{ pv(q_m^2+q_e^2) } \nonumber \\
     & \times \ev{ {W}_d^{(q_m,-q_e)}({\gamma}) }_{\tilde{\beta}, \tilde{\theta}}^p  ,
\end{align}
and
\begin{align}
    \ev{ \tilde{W}_d^{(q_e,q_m)}(F({\gamma})) }_{\beta, \theta}^p
    = &\exp(iu\pi \frac{q^2}{\tau} + 2\pi^2 u ( \beta q_m^2 -\tilde{\beta} q_e^2) )  (-1)^{ pv(q_m^2+q_e^2) } \nonumber \\
     & \times \ev{ \tilde{W}_d^{(q_m,-q_e)}(F({\gamma})) }_{\tilde{\beta}, \tilde{\theta}}^p  ,
\end{align}
where $q$ is a complex charge given by
\begin{align}
    q= q_e + \tau q_m.
\end{align}
The $\mathcal{S}$ acts on $(q_e, q_m)$ as
\begin{align}
    \mqty( q_e\\ q_m) \mapsto \mqty( q_m \\ -q_e)= \mqty(0 & 1 \\ -1 & 0)  \mqty( q_e\\ q_m) \label{eq: S for charge}.
\end{align}
\end{lemma}

\begin{remark}[Modular transformation in $(q,\tau)$]
We define the conjugate of $M \in \mathrm{SL}(2,\mathbb{Z})$ by
\begin{align}
    M^\ast=  \mqty( a & -b \\ -c & d ).
\end{align}
Then Eqs.~\eqref{eq: T for charge} and \eqref{eq: S for charge} indicate that the $\mathrm{SL}(2,\mathbb{Z})$-transformation for $(q_e, q_m)$ is given by
\begin{align}
    \mqty( q_e \\q_m ) \mapsto \mqty( q_e^\prime \\q_m^\prime ) =M^\ast\mqty( q_e \\q_m ) =\mqty( a q_e -b q_m \\ -c q_e + dq_m  ). 
\end{align}

It is useful to consider the complex structure $\tau$ and the complex charge $q$, simultaneously. Transforming $\tau$ and $q$, we get
\begin{align}
\tau\mapsto &\tau^\prime =M\tau=\frac{a\tau +b}{ c\tau +d },\\
    q\mapsto &q_e^\prime + \tau^\prime q_m^\prime = \frac{q}{c\tau+d} 
\end{align}
Then, we can express the transformation by
\begin{align}
   (q,\tau)\mapsto  M(q,\tau)= &\qty(   \frac{q}{c\tau+d},\frac{a\tau +b}{c\tau +d} ).
\end{align}

\end{remark}

The Lemmas~\ref{lem: T} and \ref{lem: S} lead to our main theorem.
\begin{theorem}[$\mathrm{SL}(2,\mathbb{Z})$-structure] \label{thm: SL(2,Z)}
We define
\begin{align}
    F\mqty[\alpha_1 \\ \alpha_2](q,\tau)= &  \exp( 2\pi^2 u  \beta q_m^2 ) (-1)^{2\alpha_1 vq_e^2} \ev{ {W}_d^{(q_e,q_m)}({\gamma}) }_{\beta, \theta}^{p=2\alpha_2} \nonumber \\ 
   =& \exp( 2\pi^2 u  \beta q_m^2 ) (-1)^{2\alpha_1 vq_e^2 +2\alpha_2 vq_m^2 } \ev{ {W}_d^{(q_e,q_m)}({\gamma}) }_{\beta, \theta}^{p=0} \label{eq: def of F}
\end{align}
for $\alpha_1,~\alpha_2 \in \frac{1}{2}\mathbb{Z}$. It transforms under the $\mathcal{S}$ and $\mathcal{T}$-transformations as follows,
\begin{align}
 F\mqty[\alpha_1 \\ \alpha_2](S( q,\tau))=F\mqty[ \alpha_1 \\\alpha_2  ] (\frac{q}{\tau},-\frac{1}{\tau})=\exp( iu \pi\frac{q^2}{\tau}  )F\mqty[ \alpha_2 \\ -\alpha_1  ](q,\tau),\\
 F\mqty[\alpha_1 \\ \alpha_2](T( q,\tau))=   F\mqty[ \alpha_1 \\\alpha_2  ] (q,\tau+1)= F\mqty[ \alpha_1 \\  \alpha_1 +\alpha_2 +1/2 ] (q,\tau).
\end{align}
Note that if $v$ is even, the four kind of functions are the same.
\end{theorem}

The quantity $u$ can be interpreted as an obstruction to constructing the $\mathcal{S}$-duality of quantum electrodynamics \cite{Anosova:2022cjm, Anosova:2022yqx}. Since the definition of $u$ involves a non-local operator, it is a non-trivial problem whether we can remove $u$ by adding a local counterterm. On the other hand, $v$ may change the statistics of Wilson loops from the analogy of non-spin Maxwell theories \cite{Ang:2019txy,Kan:2024fuu}. We will discuss this later.

\subsection{$\mathcal{T}$-transformation and Witten Effect}


We start from the expectation value of the dyonic Wilson loop in Eq.~\eqref{eq: ev of dyonic loop} and prove Lemma~\ref{lem: T}. We present the proof only for $\ev{ W_d}$, since the argument for $\ev{ \tilde{W}_d}$ is analogous.

Integrating $A^m$ generates a new constraint,
\begin{align}
     \epsilon_{\mu \nu \rho \sigma} (dn)_{x+ \hat{\mu}, \nu \rho \sigma}=- q_m \dot{\gamma}_{x, \mu}\label{eq: monopole constraint}.
\end{align}
Note that the winding number of loops that satisfy this constraint is always zero. In other words, if $\gamma$ is not contractible, the expectation value of the Wilson loop is zero. By shifting $\theta$ by $2\pi$, we have
\begin{align}
     \ev{ W_d^{(q_e,q_m)}({\gamma}) }_{\beta, \theta+2\pi}^p=  \ev{ W_d^{(q_e,q_m)}({\gamma}) \exp(- i 2\pi Q[ F^e ] )  }_{\beta, \theta}^p
\end{align}
In the presence of the magnetic Wilson loop, the topological charge is no longer an integer. The additive phase factor is given by
\begin{align}
    2\pi Q[ F^e ]=\sum_x A^e \cup dn + \pi \mathfrak{P}(n)= -q_m \dot{\gamma}\cdot A^e+\pi \mathfrak{P}(n).
\end{align}

In the presence of the monopole, the Pontryagin square can take odd values. We decompose the $2$-form $n$ into two parts:
\begin{align}
    n= q_m l +n^\prime,
\end{align}
where $n^\prime$ is an integer-valued $2$-forms with $dn^\prime=0$. 
We then get the following,
\begin{align}
\mathfrak{P}(n)
=& q_m^2\mathfrak{P}(l) +2q_m (n^\prime \cup l)+ \mathfrak{P}(n^\prime) =q_m^2v +2\mathbb{Z}.
\end{align}
Since the last term is even, it can be neglected in the exponent. That is, the additive phase factor is only determined by the contour $\gamma$ and the magnetic charge $q_m$. 

Then we obtain
\begin{align}
     \ev{ W_d^{(q_e,q_m)}({\gamma}) }_{\beta, \theta+2\pi}^p=&  \ev{ W_d^{(q_e+q_m,q_m)}({\gamma})  }_{\beta, \theta}^{p+1}\nonumber \\
     =& \ev{ W_d^{(q_e+q_m,q_m)}({\gamma})  }_{\beta, \theta}^{p}(-1)^{vq_m^2},\label{eq: T duality dyon}
\end{align}
and find that the magnetic Wilson loop dresses the electric Wilson loop with $q_m$ by the $\mathcal{T}$-transformation, which is the Witten effect. Remarkably, the electric Wilson loop is not smeared. However, the discrete theta angle is shifted by $1$ due to the magnetic Wilson loop. As a result, the periodicity of $\theta$ is essentially $4\pi$ rather than $2\pi$. This phenomenon is also found in lattice Chern-Simons theories in the modified Villain formulation \cite{Jacobson:2023cmr,Jacobson:2024hov}.

%
\subsection{Poisson summation}

Next, we tackle Lemma~\ref{lem: P}. 
The argument is analogous to the previous section, but we need to take care of the additive phase factor due to the monopole constraint Eq.~\eqref{eq: monopole constraint}. Then the topological charge can be expressed as
\begin{align}
    Q[F^e]= Q_{NL}[F^e]+ q_m^2 \frac{u}{2}
\end{align}
under the constraint. 
Using the Poisson summation formula, the expectation value of the dyonic Wilson loop is given by
\begin{align}
    \ev{ W_d^{(q_e,q_m)}({\gamma}) }_{\beta, \theta}^p
    =& \frac{1}{Z[\beta ,\theta]} \int DA^e DA^m \sum_{\{n\}} \nonumber \exp( -\frac{\beta}{2}  (F^e)^2  - i \theta Q_{NL}[ F^e ] + i\frac{1}{2\pi} \partial A^m F^e  ) \\
   &\times \exp(i\sum_{x,\mu} \dot{\gamma}_{x,\mu} \qty(q_e A^e_{x,\mu}  +  q_m \tilde{A}^m_{F(x),\mu} ) -i\theta \frac{u}{2} q_m^2  -ip \pi q_m^2 v)  \nonumber \\
   =&\frac{1}{Z[\tilde{\beta} ,\tilde{\theta}]} \int D\tilde{A}^e D\tilde{A}^m \sum_{\{m\}} \nonumber \exp( -\frac{\tilde{\beta}}{2}  (\tilde{F}^m)^2  - i \tilde{\theta} \tilde{Q}_{NL}[ \tilde{F}^m ] + i\frac{1}{2\pi} \partial \tilde{A}^e \tilde{F}^m  ) \nonumber \\
   &\times \exp(i\sum_{ \tilde{x},\mu} \dot{ \tilde{\gamma}}_{\tilde{x},\mu} \qty(q_e A^e_{F^{-1}(\tilde{x}),\mu}  +  q_m \tilde{A}^m_{\tilde{x},\mu} ) -i\theta \frac{u}{2} q_m^2  -ip \pi q_m^2 v ) . 
\end{align}
with $\dot{\tilde{\gamma}}_{F(x),\mu} = \dot{\gamma}_{x,\mu}$. By the definition of $\tilde{A}^e$, we have
\begin{align}
    \sum_{ \tilde{x},\mu} \dot{ \tilde{\gamma}}_{\tilde{x},\mu} \qty(q_e A^e_{F^{-1}(\tilde{x}),\mu}  +  q_m \tilde{A}^m_{\tilde{x},\mu} )=\sum_x \qty( q_m \dot{\tilde{\gamma}}  \tilde{A}^m - q_e \tilde{A}^e\cup \dot{\tilde{\gamma}})
\end{align}
Then, the charge $(q_e,q_m)$ of the dyonic operator changes to $(q_m,-q_e)$. Note that the dyonic particle on the dual lattice is formed by the electric charge at $\tilde{x}$ and the magnetic charge at $F^{-1}(\tilde{x})=\tilde{x}-\frac{1}{2}\hat{s}$ as shown in the right of Fig.~\ref{fig: dyonicWilsonloop}. By integrating $\tilde{A}^e$, we also find the monopole constraint on the dual lattice,
\begin{align}
     \epsilon_{\mu \nu \rho \sigma} (d\tilde{m})_{\tilde{x}-\hat{s}+ \hat{\mu}, \nu \rho \sigma}= q_e \dot{\tilde{\gamma}}_{\tilde{x}, \mu}\label{eq: monopole constraint on dual}.
\end{align}
By substituting it into Eqs.~\eqref{eq: tilde Q non local expression} and \eqref{eq: Pontryagin square}, we find
\begin{align}
    \tilde{Q}[\tilde{F}^m]=&\tilde{Q}_{NL}[\tilde{F}^m]- q_e^2 u\frac{1}{2} 
\end{align}
and 
\begin{align}
    \mathfrak{P}(\tilde{m})= q_e^2 v. 
\end{align}

Therefore, we obtain
\begin{align}
    \ev{ W_d^{(q_e,q_m)}({\gamma}) }_{\beta, \theta}^p
    =&\ev{ \tilde{W}_d^{(q_m,-q_e)}(F(\gamma)) }_{\tilde{\beta}, \tilde{\theta}}^p \exp( -i\theta u\frac{1}{2} q_m^2 -i\tilde{\theta} u\frac{1}{2} q_e^2 -ip\pi v(q_m^2+ q_e^2)  ) \label{eq: S duality dyon}.
\end{align}

This equation establishes an identity between expectation values in the dual theories. In particular, simply applying the Poisson summation formula does not yield the $\mathcal{S}$-transformation. We denote this map by $\mathcal{P}$ and distinguish it from $\mathcal{S}$.

If we use the Poisson summation formula again, we get
\begin{align}
    \ev{ W_d^{(q_e,q_m)}({\gamma}) }_{\beta, \theta}^p
    =\ev{ W_d^{(-q_e,-q_m)}({\gamma}) }_{\beta, \theta}^p.
\end{align}
This equation is also proved by flipping $A^e\to -A^e,~A^m\to -A^m$ and $n\to -n$ in the path integral. 

\subsection{Flipping framings and $\mathcal{S}$-transformation}
In this subsection, we prove Lemmas~\ref{lem: framing} and \ref{lem: S}. To do this, it is convenient to introduce 
\begin{align}
    f^p( q_e,q_m,\tau)&=  \ev{ W_d^{(q_e,q_m)}({\gamma}) }_{\beta, \theta}^p \exp( +i\theta u\frac{1}{2} q_m^2 ), \\
    \tilde{f}^p(q_e,q_m,\tau)&=  \ev{ \tilde{W}_d^{(q_e,q_m)}(F({\gamma})) }_{\beta, \theta}^p \exp( -i\theta u\frac{1}{2} q_m^2 ).
\end{align}
$f^p$ (resp. $\tilde{f}^p$) is equivalent to the expectation value of $W_d$ by replacing $Q$ (resp. $\tilde{Q}$) by $Q_{NL}$ (resp. $Q_{NL}$). From Lemmas~\ref{lem: T} and \ref{lem: P}, we find
\begin{align}
    \mathcal{T}: f^p( q_e-q_m,q_m,\tau+1) &= f^p( q_e , q_m,\tau) \exp( i \pi (u+v)q_m^2 ), \\
    \mathcal{T}: \tilde{f}^p( q_e-q_m,q_m,\tau+1)&= \tilde{f}^p( q_e , q_m, \tau)\exp( i \pi (-u+v)q_m^2 ),
\end{align}
and
\begin{align}
\mathcal{P}: \tilde{f}^p( q_m,-q_e, -1/\tau)&= (-1)^{ pv(q_m^2+q_e^2) }f^p(q_e,q_m, \tau ).
\end{align}

Alternating $\mathcal{P}$ and $\mathcal{T}$-three times, we have
\begin{align}
    f^p( q_e,q_m,\tau)=&f^p( -q_e,-q_m,\tau) \nonumber \\
    =& \exp( i\pi pv (q_e^2+ q_m^2  ) + i\pi (-u+v)q_e^2 ) \tilde{f}( -q_e +q_m, -q_e ,STST \tau ) \nonumber \\
    =& \cdots \nonumber \\
    =&\exp( i\pi p v (q_e^2+ q_m^2 + q_e^2 +(-q_e+q_m)^2 +(-q_e+q_m)^2+q_m^2   ))  \nonumber \\
   & \times\exp( i\pi \qty[ (-u+v)q_e^2 + (u+v) (-q_e^2+q_m^2)+(-u+v) q_m^2    ] ) \tilde{f}( q_e,q_m ,\tau ) \nonumber \\
   =& \exp( i\pi u (-2 q_eq_m) ) \tilde{f}^p( q_e,q_m ,\tau ) \label{eq: PTPTPT}.
\end{align}
This leads to Lemma~\ref{lem: framing},
\begin{align}
    \ev{ \tilde{W}_d^{(q_e,q_m)}(F{\gamma}) }_{\beta, \theta}^p
    =&\ev{ {W}_d^{(q_e,q_m)}({\gamma}) }_{\beta, \theta}^p\exp( i 2\pi u(q_e+\frac{\theta}{2\pi}q_m )q_m  )
\end{align}
The phase factor can be interpreted as the Aharonov-Bohm phase between the electric charge $q_e+ \frac{\theta}{2\pi}q_m$ and the magnetic charge $q_m$ \cite{Jackiw:1976xx, Goldhaber:1976dp, Hasenfratz:1976gr}. Note that $u$ can take non-integer values due to non-trivial self-linking of $\gamma$ \footnote{From the analogy with lattice Chern–Simons theory \cite{Jacobson:2023cmr, Jacobson:2024hov, Chen:2019mjw, Xu:2024hyo, Peng:2025nfa, Eliezer:1991qh}, this phase factor may be interpreted as related to the framing anomaly. However, we leave a detailed investigation of the physical meaning of $u$ for future work.}. 

Applying the Poisson summation again, we get
\begin{align}
     f^p(q_e,q_m, \tau   )=\exp( -i\pi u (2 q_eq_m) ) f^p( q_m,-q_e,-1/\tau)(-1)^{ pv(q_m^2+q_e^2) }.
\end{align}
We can rewrite it in terms of the expectation value,
\begin{align}
    \ev{ W_d^{(q_e,q_m)}({\gamma}) }_{\beta, \theta}^p= \exp( -i u \pi \qty[ \frac{\theta }{2\pi } q_m^2 + 2 q_eq_m-\frac{\tilde{\theta} }{2\pi } q_e^2   ])  (-1)^{ pv(q_m^2+q_e^2) } \nonumber \\
    \times \ev{ {W}_d^{(q_m,-q_e)}({\gamma}) }_{\tilde{\beta}, \tilde{\theta}}^p   \nonumber \\
    = \exp(-iu\pi \frac{q^2}{\tau} - 2\pi^2 u ( \beta q_m^2 -\tilde{\beta} q_e^2) )(-1)^{ pv(q_m^2+q_e^2) }\nonumber \\
      \times  \ev{ {W}_d^{(q_m,-q_e)}({\gamma}) }_{\tilde{\beta}, \tilde{\theta}}^p  .
\end{align}
Then Lemma~\ref{lem: S} holds.

\subsection{$\mathrm{SL}(2,\mathbb{Z})$-structure}
Finally, we prove the main theorem. Recalling the definition \eqref{eq: def of F} and applying Lemma~\ref{lem: T}, we obtain
\begin{align}
F\mqty[\alpha_1 \\ \alpha_2](T( q,\tau))&=F\mqty[ \alpha_1 \\\alpha_2  ] \qty(q,\tau+1)=F\mqty[ \alpha_1 \\\alpha_2  ] \qty((q_e-q_m) + \tau q_e,\tau+1)  \nonumber \\
&=\exp( 2\pi^2 u  \beta q_m^2 ) (-1)^{2\alpha_1 v(q_e-q_m)^2 +2\alpha_2 vq_m^2 } \ev{ {W}_d^{(q_e-q_m,q_e)}({\gamma}) }_{\beta, \theta+2\pi}^{p=0} \nonumber \\
&= (-1)^{2\alpha_1 v (q_e^2+q_m^2) +2\alpha_2 vq_m^2 +vq_m^2 } \exp( 2\pi^2 u  \beta q_m^2 ) \ev{ {W}_d^{(q_e,q_m)}({\gamma}) }_{\beta, \theta}^{p=0} \nonumber \\
&=F\mqty[\alpha_1 \\ \alpha_1+ \alpha_2+ 1/2]( q,\tau).
\end{align}
Also using Lemma~\ref{lem: S}, we find
\begin{align}
F\mqty[\alpha_1 \\ \alpha_2](S( q,\tau))&=F\mqty[ \alpha_1 \\\alpha_2  ] \qty(\frac{q}{\tau}, -\frac{1}{\tau})=F\mqty[ \alpha_1 \\\alpha_2  ] \qty(q_m + \qty(-\frac{1}{\tau})(-q_e),-\frac{1}{\tau},)  \nonumber \\
&=\exp( 2\pi^2 u  \tilde{\beta} q_e^2 ) (-1)^{2\alpha_1 vq_m^2 +2\alpha_2 vq_e^2 } \ev{ {W}_d^{(q_m,-q_e)}({\gamma}) }_{\tilde{\beta}, \tilde{\theta}}^{p=0} \nonumber \\
&=\exp( iu\pi \frac{q^2}{\tau} )  (-1)^{-2\alpha_1 vq_m^2 +2\alpha_2 vq_e^2 } \exp( 2\pi^2 u  \beta q_m^2 ) \ev{ {W}_d^{(q_e,q_m)}({\gamma}) }_{\beta, \theta}^{p=0} \nonumber \\
&=\exp( iu\pi \frac{q^2}{\tau} )F\mqty[\alpha_2 \\ -\alpha_1]( q,\tau).
\end{align}

The exponential function defines a factor of automorphy of $\mathrm{SL}(2,\mathbb{Z})$, given by
\begin{align}
    e(M, (q,\tau))= \exp\left( iu\pi \frac{cq^2}{c\tau+d} \right),
\end{align}
which satisfies
\begin{align}
    e(M_1 M_2, (q,\tau))= e(M_1 ,M_2 (q,\tau))\, e(M_2, (q,\tau)).
\end{align}
That is, the transformation law is consistent with the group action.

\begin{figure}
    \centering
    \includegraphics[width=\linewidth,bb=0 0 964 143]{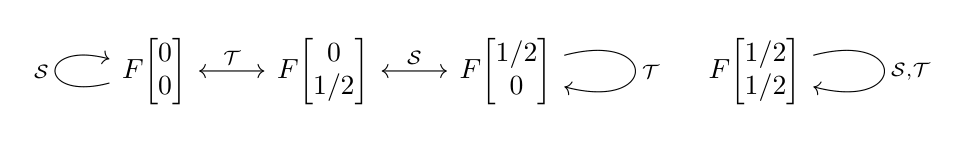}
    \caption{$\mathrm{SL}(2,\mathbb{Z})$ orbit of the expectation value of the dyonic Wilson loops. ${{F}\mqty[0 \\ 0]}$ and $ {{F}\mqty[0 \\ 1/2]} $ are related to the expectation value of $\ev{W_d}^{p=0}$ and $\ev{W_d}^{p=1}$. $\mathcal{S}$ and $\mathcal{T}$ are generators of $\mathrm{SL}(2,\mathbb{Z})$. }
    \label{fig: Modular rep of Wilson loops}
\end{figure}

The group $\mathrm{SL}(2,\mathbb{Z})$ also acts on $(\alpha_1,\alpha_2)$ as an element of $\mathbb{Z}_2 \times \mathbb{Z}_2$. This action is the same as that of the theta functions, and reduces modulo $2$ to $\mathrm{SL}(2,\mathbb{Z}_2)$, which is isomorphic to the symmetric group $\mathfrak{S}_3$. Under this action, we find two orbits: one is the trivial orbit
\begin{align}
    (\alpha_1,\alpha_2)=(1/2,1/2),
\end{align}
while the remaining three elements are permuted among themselves as shown in Fig.~\ref{fig: Modular rep of Wilson loops},
\begin{align}
    (\alpha_1,\alpha_2)=(0,0),~ (0,1/2),~ (1/2,0).
\end{align}

Such a $\mathrm{SL}(2,\mathbb{Z})$-structure is also found in the non-spin Maxwell theory \cite{Ang:2019txy, Kan:2024fuu}, where the second Stiefel-Whitney class $w_2$ is non-trivial, and the topological charge becomes a half integer. $w_2$ can be interpreted as a background $2$-form gauge field of electric and magnetic $1$-form symmetry. There are four ways to turn on the background gauge field, which corresponds to the four kinds of the dyonic Wilson loop, $W_\text{b} T_\text{b},~W_\text{b} T_\text{f},~ W_\text{f}T_\text{b}$ and $W_\text{f}T_\text{f}$. The subscripts mean the bosonic and fermionic statistics of the Wilson and 't Hooft loops. For example, the fermionic Wilson loop is defined by
\begin{align}
    W_\text{f}^{q_e}(\gamma)= (-1)^{q_e \int_\Sigma w_2 } \exp( i q_e\oint_\gamma A^e ),
\end{align}
with $\partial \Sigma= \gamma$.

Compared to our case, the magnetic Wilson loop renders the Pontryagin square half-integer valued, and the solution of the monopole constraint Eq.~\eqref{eq: monopole constraint} plays the role of $w_2$. That is,
\begin{align}
    v \longleftrightarrow \int_\Sigma w_2.
\end{align}

From the analogy of non-spin Maxwell theories, we can define the angular momentum of the dyonic Wilson loop as
\begin{align}
    J= (2q_e  \alpha_1+2 q_m \alpha_2 + q_e q_m)v \pmod{2}.
\end{align}
$\mathcal{S}$ and $\mathcal{T}$ commute with $(-1)^J$ whose eigenvalue is related to the statistics of dyonic matters \cite{Jackiw:1976xx, Goldhaber:1976dp, Hasenfratz:1976gr}. Then, $\alpha_1$ and $\alpha_2$ can be interpreted as the spin of electric and magnetic Wilson loops\footnote{We need more discussion to identify that $\alpha_1$ and $\alpha_2$ induce the spin to the Wilson loops since we have not demonstrated the behavior of Wilson loops under $2\pi$-rotation. }.


Our results are consistent with lattice Chern–Simons theory in the modified Villain formulation \cite{Jacobson:2023cmr, Jacobson:2024hov}. The action requires a $\mathbb{Z}$ 1-form symmetry, and its quantization level must be even. This is related to the $4\pi$-periodicity of $\theta$, rather than $2\pi$. In contrast, the odd-level case corresponds to a spin theory, in which the Wilson loop acquires fermionic statistics \cite{Gaiotto:2015zta,Bhardwaj:2016clt, Belov:2005ze, Dijkgraaf:1989pz,Jacobson:2024hov, Xu:2024hyo}. Our observed $\mathrm{SL}(2,\mathbb{Z})$-structure may reflect the statistical transmutation.

\section{Conclusion and Discussion}
\label{sec: conclusion}

In this paper, we discussed the duality of the four dimensional $U(1)$ Maxwell theory. Remarkably, the ultra-local action possesses the exact $\mathrm{SL}(2,\mathbb{Z})$-duality. We also clarified the $\mathrm{SL}(2,\mathbb{Z})$-structure of the dyonic Wilson loops. The structure is closer to that found in the non-spin Maxwell theory rather than Maxwell theory on spin manifolds. That is because the $\theta$-term can take half-integer values in the presence of the magnetic Wilson loop.

As discussed in Ref.~\cite{Anosova:2022cjm}, the $\mathrm{SL}(2,\mathbb{Z})$ structure may provide a possible strategy for mitigating the complex action problem associated with the $\theta$ term by mapping the theory to a duality frame. If the theory is mapped to a duality frame with $\theta=0$, the $\theta$ term is eliminated. In this duality frame, the non-ultra-locality present in Ref.~\cite{Anosova:2022cjm} also disappears, and Monte Carlo simulations can be performed using an ultra-local action. Therefore, the present ultra-local formulation does not directly improve the efficiency of Monte Carlo simulations. Rather, its main advantage is that it establishes the relation between different duality frames exactly at finite lattice spacing, providing a rigorous foundation for applying duality transformations to the study of the complex action problem.



We are interested in $\mathrm{SL}(2,\mathbb{Z})$-duality in the canonical formalism. The Hamiltonian of Maxwell theory in the modified Villain formulation is discussed in \cite{Fazza:2022fss}, where the $\mathcal{S}$-duality was also analyzed. The $\mathcal{S}$-transformation can be described by a canonical transformation \cite{Lozano:1995aq} and exchanges electric and magnetic fields. 
On the other hand, the $\mathcal{T}$-transformation is defined in terms of a Chern-Simons term, which is the boundary term of the topological charge. Once the $\mathcal{S}$- and $\mathcal{T}$-transformations have been defined, we can clarify the resulting $\mathrm{SL}(2,\mathbb{Z})$-structure. We are currently working on this problem.



Our model is expected to reformulate the Cardy-Rabinovici (CR) model \cite{Cardy:1981fd, Cardy:1981qy,Honda:2020txe,Hayashi:2022fkw, Katayama:2025pmz}. The CR model is a toy model of a lattice $U(1)$ gauge theory, which exhibits the complicated phase diagram, including the Coulomb, Higgs, and confinement phases. Adding the Higgs fields to our theory, the effective action is expected to be the CR model with some modifications. Furthermore, our method makes it possible to construct the non-invertible defect \cite{Choi:2022zal, Choi:2021kmx} on lattice spaces without a formal continuum description \cite{Hayashi:2022fkw}. 



In four dimensions, in addition to Abelian gauge theories, it is known that $\mathcal{N}=4$ super Yang--Mills (SYM) theory possesses an exact $\mathrm{SL}(2,\mathbb{Z})$-duality. $\mathcal{N}=4$ SYM also provides a prototypical example of the AdS/CFT correspondence, being dual to type IIB string theory on $AdS_5 \times S^5$.
Therefore, understanding its dynamics is expected to shed light on nonperturbative aspects of quantum gravity. Lattice formulations of $\mathcal{N}=4$ SYM have been proposed in, e.g., \cite{Catterall:2005fd, Catterall:2009it, Giedt:2016yfw, Sugino:2003yb, Kaplan:2002wv}.
In particular, in \cite{Giedt:2016yfw}, $\mathcal{S}$-duality has been tested in the Coulomb branch. It would be interesting if we establish the $\mathrm{SL}(2,\mathbb{Z})$-duality on the lattice SYM model as the lattice Maxwell theory.




Although our method is based on the hypercubic lattice, it is possible to extend all of the features investigated in this work to general triangular lattice spaces. Although the Poisson summation maps to the dual lattice, it returns to the original lattice by the prescription presented in the main text. The important part for the $\mathrm{SL}(2,\mathbb{Z})$ is the harmonic part of the $2$-form, which is directly related to the topology of the manifold. According to the continuum theory \cite{Witten:1995gf}, there is a mixed 't Hooft anomaly between its background gravity and the $\mathrm{SL}(2,\mathbb{Z})$-duality when its signature is non-trivial. The famous examples are $\mathbb{C}P^2$ and $K3$, whose triangulations are presented in \cite{Kuhnel1983} and \cite{CASELLA2001753}. If it is possible to extend lattice Maxwell theory to their triangulations, one may be able to observe the gravitational anomaly discussed in \cite{Honda:2020txe,Hayashi:2022fkw} on the lattice.


%

\acknowledgments

We would like to thank Y. Furukawa, Y. Ikeda, Y. Murakami, H. Wada for valuable and enlightening discussions. S.A. is supported by RIKEN Special Postdoctoral Researchers Program and JSPS KAKENHI Grant Number 25K17382. T.T. is supported by JST SPRING, Grant Number JPMJSP2108.

\appendix

\section{Differential Forms on the Lattice}
\label{App: Differential Form}
In this appendix, we review differential forms on the lattice following \cite{Anosova:2022cjm}. In the main text, we have mainly considered a four-dimensional lattice, however, here we consider a lattice of general dimension. 

We consider a $d$-dimensional hypercubic lattice $\Lambda$ on the torus, and denote lattice sites by $x = (x_{1},x_{2},\cdots,x_{d})$.
We define $r$-cells in the lattice as oriented elementary hypercubes of dimension $r$.
In particular, $0$-cells correspond to lattice sites, $1$-cells to links, and $2$-cells to plaquettes. More generally, an $r$-cell is specified by a lattice site $x$ together with an ordered set of directions $\mu_{1}\mu_{2}\cdots\mu_{r}$ ($\mu_{1}<\mu_{2}<\cdots<\mu_{r}$), and is denoted by $(x,\mu_{1}\mu_{2}\cdots\mu_{r})$. 

The $r$-forms are defined on the $r$-cells $(x,\mu_{1}\mu_{2}\cdots\mu_{r})$ of the lattice, whose value is assigned by $\alpha_{x,\mu_{1}\mu_{2}\cdots\mu_{r}}$.
We assume that the field is antisymmetric with respect to permutations of the direction indices, 
$\alpha_{x,\mu_{1}\cdots\mu_{i}\cdots\mu_{j}\cdots\mu_{r}}=-\alpha_{x,\mu_{1}\cdots\mu_{j}\cdots\mu_{i}\cdots\mu_{r}}$. We denote the set of $K$-valued $r$-form by $C^{r}(\Lambda, K)  $ with $K= \mathbb{Z},\mathbb
{R}$, or $\mathbb{C}$.

We now define two types of discrete differential operators. The exterior derivative 
$d^{(r)}:C^{r}(\Lambda, K) \to C^{r+1}(\Lambda, K) $ is an operator that maps 
$r$-forms to $(r+1)$-forms, while the boundary operator $\partial^{(r)}:C^{r}(\Lambda, K)\to C^{r-1}(\Lambda, K)$ maps $r$-forms to $(r-1)$-forms.
They are defined as follows
\begin{align}
&(d^{(r)}\alpha)_{x,\mu_{1}\mu_{2}\cdots\mu_{r}} = \sum_{j=1}^{r}(-1)^{j+1}[\alpha_{x+\hat{\mu_{j}},\mu_{1}\cdots\mathring{\mu}_j \cdots \mu_{r}}-\alpha_{x,\mu_{1}\cdots\mathring{\mu}_j \cdots \mu_{r}}],\\
&(\partial^{(r)} \alpha)_{x,\mu_{1}\mu_{2}\cdots\mu_{r}} = \sum_{\nu=1}^{d}[\alpha_{x,\mu_{1}\cdots\mu_{r},\nu}-\alpha_{x-\nu,\mu_{1}\cdots \mu_{r}\nu}].
\end{align}
Here, the notation $\mathring{\mu}_{j}$ means that the index $\mu_{j}$ is excluded from the set of indices. The superscript that refers to its domain is omitted when there is no risk of confusion. These two operators satisfy the nilpotency conditions $d^{2}=0$ and $\partial^{2}=0$.
In momentum space, exterior derivative and boundary operator act as
\begin{align}
&(d\alpha)'_{\mu_{1}\mu_{2}\cdots\mu_{r}}(p) = \sum_{j=1}^{r}(-1)^{j+1}f_{\mu_{j}}\alpha'_{\mu_{1}\cdots\mathring{\mu}_j \cdots \mu_{r}}(p),\\
&(\partial \alpha) '_{\mu_{1}\mu_{2}\cdots\mu_{r}}(p)=\sum_{\nu=1}^{d}-f^{*}_{\nu}\alpha'_{\mu_{1}\cdots \mu_{r}\nu}(p),
\end{align}
where $f_\mu(p)= e^{ip_\mu} -1$ and $ \alpha_{x,\mu_{1}\mu_{2}\cdots\mu_{r} }= \sum_{p } \frac{e^{ipx}}{ \sqrt{N^{d}}} {\alpha}^\prime_{ \mu_{1}\mu_{2}\cdots\mu_{r} } (p)$.
Moreover, the following partial integration formula holds.
\begin{align}
\sum_{x}\sum_{\mu_{1}<\cdots<\mu_{r}}\alpha_{x,\mu_{1}\mu_{2}\cdots\mu_{r}} (d\beta)_{x,\mu_{1}\mu_{2}\cdots\mu_{r}}=(-1)^{r}\sum_{x}\sum_{\mu_{1}<\cdots<\mu_{r-1}}(\partial\alpha)_{x,\mu_{1}\mu_{2}\cdots\mu_{r-1}} \beta_{x,\mu_{1}\mu_{2}\cdots\mu_{r-1}}
\end{align}
where $\alpha$ is an $r$-form and $\beta$ is an $r-1$-form. This formula can be shown as follows.
\begin{align}
&\sum_{x}\sum_{\mu_{1}<\cdots<\mu_{r}}\alpha_{x,\mu_{1}\mu_{2}\cdots\mu_{r}} (d\beta)_{x,\mu_{1}\mu_{2}\cdots\mu_{r}} \nonumber\\
&= \sum_{x}\sum_{\mu_{1}<\cdots<\mu_{r}}\sum_{j=1}^{r}(-1)^{j+1}\alpha_{x,\mu_{1}\mu_{2}\cdots\mu_{r}} [\beta_{x+\hat{\mu_{j}},\mu_{1}\cdots\mathring{\mu}_j \cdots \mu_{r}}-\beta_{x,\mu_{1}\cdots\mathring{\mu}_j \cdots \mu_{r}}] \nonumber \\
&=\sum_{x}\sum_{\mu_{1}<\cdots<\mu_{r}}\sum_{j=1}^{r}(-1)^{j+1}[\alpha_{x-\hat{\mu_{j}},\mu_{1}\cdots \mu_{r}}-\alpha_{x,\mu_{1}\cdots \mu_{r}}]\beta_{x,\mu_{1}\mu_{2}\cdots\mathring{\mu}_j\cdots \mu_{r}} \nonumber \\
&=\sum_{x}\sum_{\mu_{1}<\cdots<\mu_{r-1}}\sum_{\nu=1}^{d}(-1)^{r}[\alpha_{x-\nu,\mu_{1}\cdots \mu_{r-1},\nu}-\alpha_{x,\mu_{1}\cdots \mu_{r-1},\nu}]\beta_{x,\mu_{1}\mu_{2}\cdots \mu_{r-1}}\nonumber \\
&=(-1)^{r}\sum_{x}\sum_{\mu_{1}<\cdots<\mu_{r-1}}(\partial\alpha)_{x,\mu_{1}\mu_{2}\cdots\mu_{r-1}} \beta_{x,\mu_{1}\mu_{2}\cdots\mu_{r-1}}
\end{align}

In the analysis of the non-local $\theta$-term, the Hodge decomposition played a crucial role. The Hodge decomposition asserts that any $\mathbb{C}$-valued $r$-form  $\alpha$ admits a unique decomposition
\begin{align}
\alpha_{x,\mu_{1}\mu_{2}\cdots\mu_{r}} = (dp)_{x,\mu_{1}\mu_{2}\cdots\mu_{r}}+(\partial q)_{x,\mu_{1}\mu_{2}\cdots\mu_{r}}+h_{x,\mu_{1}\mu_{2}\cdots\mu_{r}}.
\end{align}
with the $(r-1)$-form field $p$, the $(r+1)$-form field $q$ and the $r$-form field $h$. That is, the set $C^{r}(\Lambda, \mathbb{C})$ can be expressed by
\begin{align}
    C^{r}(\Lambda, \mathbb{C})= \text{Im}( d^{(r-1)}  )\oplus \text{Im}( \partial^{(r+1)}  ) \oplus H^r(\Lambda, \mathbb{C}),
\end{align}
where $H^r(\Lambda,\mathbb{C}) $ denotes the set of harmonic elements, 
\begin{align}
    H^r(\Lambda,\mathbb{C})= \{ \alpha \in C^r(\Lambda,\mathbb{C})  \mid d^{(r)} \alpha =0,~\partial^{(r)}\alpha=0  \}. 
\end{align}
$p,~q$ and $h$ are expressed as
\begin{align}
&p_{x, \mu_{1}\mu_{2}\cdots\mu_{r-1}}= \sum_{p\neq 0 } \frac{e^{ipx}}{ \sqrt{N^{d}}} \frac{(-1)^{r}{(\partial \alpha)}^\prime_{ \mu_{1}\mu_{2}\cdots\mu_{r-1} } (p)}{\abs{f}^2},\\
&q_{x, \mu_{1}\mu_{2}\cdots\mu_{r+1}}= \sum_{p\neq 0 } \frac{e^{ipx}}{ \sqrt{N^{d}}} \frac{(-1)^{r-1}{(d \alpha)}^\prime_{ \mu_{1}\mu_{2}\cdots\mu_{r+1} } (p)}{\abs{f}^2},\\
&h_{x, \mu_{1}\mu_{2}\cdots\mu_{r}}=\frac{1}{\sqrt{N^{d}}}\alpha'_{\mu_{1}\mu_{2}\cdots\mu_{r}}(0).
\end{align}
Note that $\abs{f}=0$ only on $p=0$. One can define projection operators $P_{d},~P_{\partial}$, $P_{0}$ onto $\text{Im}( d^{(r-1)}  ),~\text{Im}( \partial^{(r+1)}  ) $ and $ H^r(\Lambda, \mathbb{C})$ by
\begin{align}
&(P_{d}\alpha)_{x, \mu_{1}\mu_{2}\cdots\mu_{r}}=dp= \sum_{p\neq 0 } \frac{e^{ipx}}{ \sqrt{N^{d}}} \sum_{j=1}^{r}\sum_{\mu_{r+1}=1}^{d}\frac{(-1)^{r-1}(-1)^{j+1}f_{\mu_{j}}f_{\mu_{r+1}}^{*}{\alpha}^\prime _{\mu_{1}\cdots\mathring{\mu}_j \cdots \mu_{r+1} } (p)}{\abs{f}^2},\\
&(P_{\partial}\alpha)_{x, \mu_{1}\mu_{2}\cdots\mu_{r}}=\partial q= \sum_{p\neq 0 } \frac{e^{ipx}}{ \sqrt{N^{d}}} \sum_{j=1}^{r+1}\sum_{\mu_{r+1}=1}^{d}\frac{(-1)^{r}(-1)^{j+1}f_{\mu_{j}}f_{\mu_{r+1}}^{*}{\alpha}^\prime _{\mu_{1}\cdots\mathring{\mu}_j \cdots \mu_{r+1} } (p)}{\abs{f}^2},\\
&(P_{0}\alpha)_{x, \mu_{1}\mu_{2}\cdots\mu_{r}}=h=\frac{1}{\sqrt{N^{d}}}\alpha'_{\mu_{1}\mu_{2}\cdots\mu_{r}}(0).
\end{align}
These operators satisfy $P_{d}+P_{\partial}+P_{0}=1$, which can be shown as follows
\begin{align}
&(P_{d}\alpha+P_{\partial}\alpha+P_{0}\alpha)_{x, \mu_{1}\mu_{2}\cdots\mu_{r}} \nonumber\\
&=\sum_{p\neq 0 } \frac{e^{ipx}}{ \sqrt{N^{d}}} \sum_{\mu_{r+1}=1}^{d}\frac{(-1)^{r}(-1)^{r+2}f_{\mu_{r+1}}f_{\mu_{r+1}}^{*}{\alpha}^\prime _{\mu_{1} \cdots \mu_{r} } (p)}{\abs{f}^2}+\frac{1}{\sqrt{N^{d}}}\alpha'_{\mu_{1}\mu_{2}\cdots\mu_{r}}(0) \nonumber \\
&=\sum_{p\neq 0 } \frac{e^{ipx}}{ \sqrt{N^{d}}} {{\alpha}^\prime _{\mu_{1} \cdots \mu_{r} } (p)}+\frac{1}{\sqrt{N^{d}}}\alpha'_{\mu_{1}\mu_{2}\cdots\mu_{r}}(0) \nonumber \\
&=\alpha_{x, \mu_{1}\mu_{2}\cdots\mu_{r}}.
\end{align}
It can be shown that $P_{i}^{2}=P_{i}$ for $i=d,\partial,0$. For $P_{d}$, one finds the following.
\begin{align}
(P_{d}P_{d}\alpha)_{x, \mu_{1}\mu_{2}\cdots\mu_{r}}&=\sum_{p\neq 0 } \frac{e^{ipx}}{ \sqrt{N^{d}}} \sum_{j=1}^{r}\sum_{\mu_{r+1}=1}^{d}\sum_{\substack{i=1\\i\neq j}}^{r+1}\sum_{\mu_{r+2}=1}^{d}\nonumber\\
&\quad \times  \frac{(-1)^{r-1}(-1)^{j+1}(-1)^{r-1}(-1)^{s(i,j)}f_{\mu_{j}}f_{\mu_{r+1}}^{*}f_{\mu_{i}}f_{\mu_{r+2}}^{*}{\alpha}^\prime _{\mu_{1}\cdots\mathring{\mu}_j \cdots\mathring{\mu}_i \cdots  \mu_{r+2} } (p)}{\abs{f}^4} \nonumber \\
&=\sum_{p\neq 0 } \frac{e^{ipx}}{ \sqrt{N^{d}}} \sum_{j=1}^{r}\sum_{\mu_{r+1}=1}^{d}\sum_{\mu_{r+2}=1}^{d}\nonumber\\
&\quad \times  \frac{(-1)^{r-1}(-1)^{j+1}f_{\mu_{j}}f_{\mu_{r+1}}^{*}f_{\mu_{r+1}}f_{\mu_{r+2}}^{*}{\alpha}^\prime _{\mu_{1}\cdots\mathring{\mu}_j \cdots\mathring{\mu}_{r+1}  \mu_{r+2} } (p)}{\abs{f}^4} \nonumber \\
&=\sum_{p\neq 0 } \frac{e^{ipx}}{ \sqrt{N^{d}}} \sum_{j=1}^{r}\sum_{\mu_{r+2}=1}^{d}\frac{(-1)^{r-1}(-1)^{j+1}f_{\mu_{j}}f_{\mu_{r+2}}^{*}{\alpha}^\prime _{\mu_{1}\cdots\mathring{\mu}_j \cdots\mathring{\mu}_{r+1} \mu_{r+2} } (p)}{\abs{f}^2} \nonumber \\
&=(P_{d}\alpha)_{x, \mu_{1}\mu_{2}\cdots\mu_{r}},
\end{align}
where 
\begin{equation}
  s(i,j)=
  \begin{cases}
    i & \text{if $i>j$,} \\
    i+1   & \text{if $i<j$.}
  \end{cases}
\end{equation}
In the second line, we used the fact that exchanging 
$i$ and $j$ produces a minus sign, so that only the case $i=r+1$ contributes.
In complete analogy, one can show that $P_{\partial}^{2}=P_{\partial}$ also holds.
The relation $P_{i}P_{j}=0$ for $i\neq j$ also holds.
For $P_{\partial}P_{d}$,
\begin{align}
(P_{\partial}P_{d}\alpha)_{x, \mu_{1}\mu_{2}\cdots\mu_{r}}&=\sum_{p\neq 0 } \frac{e^{ipx}}{ \sqrt{N^{d}}} \sum_{j=1}^{r}\sum_{\mu_{r+1}=1}^{d}\sum_{\substack{i=1\\i\neq j}}^{r+2}\sum_{\mu_{r+2}=1}^{d}\nonumber\\
&\quad\times  \frac{(-1)^{r-1}(-1)^{j+1}(-1)^{r}(-1)^{s(i,j)}f_{\mu_{j}}f_{\mu_{r+1}}^{*}f_{\mu_{i}}f_{\mu_{r+2}}^{*}{\alpha}^\prime _{\mu_{1}\cdots\mathring{\mu}_j \cdots\mathring{\mu}_i \cdots  \mu_{r+2} } (p)}{\abs{f}^4} \nonumber \\
&=\sum_{p\neq 0 } \frac{e^{ipx}}{ \sqrt{N^{d}}} \sum_{j=1}^{r}\sum_{\mu_{r+2}=1}^{d} \frac{(-1)^{r}(-1)^{j+1}f_{\mu_{j}}f_{\mu_{r+2}}^{*}{\alpha}^\prime _{\mu_{1}\cdots\mathring{\mu}_j \cdots\mathring{\mu}_{r+1} \mu_{r+2} } (p)}{\abs{f}^2}\nonumber\\
&\quad+\sum_{p\neq 0 } \frac{e^{ipx}}{ \sqrt{N^{d}}} \sum_{j=1}^{r}\sum_{\mu_{r+1}=1}^{d} \frac{(-1)^{r-1}(-1)^{j+1}f_{\mu_{j}}f_{\mu_{r+1}}^{*}{\alpha}^\prime _{\mu_{1}\cdots\mathring{\mu}_j \cdots \mu_{r+1}\mathring{\mu}_{r+2} } (p)}{\abs{f}^2} \nonumber \\
&=0.
\end{align}
Similarly, one can show that $P_{d}P_{\partial}=0$. 

Since the restriction of $d$ onto $\text{Im}(\partial)$ is a bijection, we can count the dimensions,
\begin{align}
    \text{dim}( C^{(r)} (\Lambda, K) ) &= {}_dC_r N^d,\\
    \text{dim}( \text{Im}(\partial^{(r+1)}) )&={}_{d-1}C_{r}(N^d-1),\\
    \text{dim}(\text{Im}( d^{(r-1)} ))&={}_{d-1}C_{r-1} (N^d-1),\\
    ~\text{dim}( H^{(r)} (\Lambda, K) )&={}_dC_r.
\end{align}

 The Hodge star is an operation that maps an $r$-form on the lattice $\Lambda$ to a $(d-r)$-form on the dual lattice $\tilde{\Lambda}$. As in the main text, the dual lattice is represented by $\tilde{\Lambda}= \{ \tilde{x}=x+ \frac{1}{2} \hat{s} \mid x \in \Lambda  \}$ where $\hat{s}=\hat{1}+\hat{2}+\cdots+\hat{d}$. The Hodge star of a $r$-form $\alpha$ is denoted by $\tilde{\alpha}$ and is given by the following relation.
\begin{align}
\alpha_{x, \mu_{1}\mu_{2}\cdots\mu_{r}}=\sum_{\nu_{r+1}<\nu_{r+2}<\cdots<\nu_{d}}\epsilon_{ \mu_{1}\mu_{2}\cdots\mu_{r}\nu_{r+1}\cdots\nu_{d}}\tilde{\alpha}_{F(x)-\hat{\nu}_{r+1}-\hat{\nu}_{r+2}\cdots-\hat{\nu}_{d},\nu_{r+1}\cdots\nu_{d}}
\end{align}
where $F(x)=x+\frac{1}{2}\hat{s}$ as in the main text.
It is also true that the exterior derivative and the boundary operator are mapped to each other under the Hodge star
\begin{align}
(\partial \alpha)_{x, \mu_{1}\mu_{2}\cdots\mu_{r}}=\sum_{\nu_{r+1}<\nu_{r+2}<\cdots<\nu_{d}}\epsilon_{ \mu_{1}\mu_{2}\cdots\mu_{r}\nu_{r+1}\cdots\nu_{d}}(d\tilde{\alpha})_{F(x)-\hat{\nu}_{r+1}-\hat{\nu}_{r+2}\cdots-\hat{\nu}_{d},\nu_{r+1}\cdots\nu_{d}}.
\end{align}
One can show this identity as follows.
\begin{align}
&(\partial \alpha)_{x, \mu_{1}\mu_{2}\cdots\mu_{r}} \nonumber \\
&=\sum_{\sigma=1}^{d}[\alpha_{x,\mu_{1}\cdots\mu_{r},\sigma}-\alpha_{x-\hat{\sigma},\mu_{1}\cdots \mu_{r}\sigma}] \nonumber\\
&=\sum_{\sigma=1}^{d}\sum_{\nu_{r+2}<\nu_{r+2}<\cdots<\nu_{d}}\epsilon_{ \mu_{1}\mu_{2}\cdots\mu_{r}\sigma\nu_{r+2}\cdots\nu_{d}}[\tilde{\alpha}_{F(x)-\nu_{r+2}\cdots-\nu_{d},\nu_{r+2}\cdots\nu_{d}}-\tilde{\alpha}_{F(x-\hat{\sigma})-\nu_{r+2}\cdots-\nu_{d},\nu_{r+2}\cdots\nu_{d}}] \nonumber\\
&=\sum_{\nu_{r+1}<\nu_{r+2}<\cdots<\nu_{d}}\epsilon_{ \mu_{1}\mu_{2}\cdots\mu_{r}\nu_{r+1}\cdots\nu_{d}}\nonumber \\
&\qquad \times\sum_{j=1}^{d-r}(-1)^{j+1}[\tilde{\alpha}_{F(x+\hat{\nu}_{r+j})-\nu_{r+1}\cdots-\nu_{d},\nu_{r+1}\cdots\mathring{\nu}_{r+j}\cdots\nu_{d}}-\tilde{\alpha}_{F(x)-\nu_{r+1}\cdots-\nu_{d},\nu_{r+1}\cdots\mathring{\nu}_{r+j}\cdots\nu_{d}}] \nonumber \\
&=\sum_{\nu_{r+1}<\nu_{r+2}<\cdots<\nu_{d}}\epsilon_{ \mu_{1}\mu_{2}\cdots\mu_{r}\nu_{r+1}\cdots\nu_{d}}(d\tilde{\alpha})_{F(x)-\nu_{r+1}\cdots-\nu_{d},\nu_{r+1}\cdots\nu_{d}}
\end{align}
In the fourth line, we set $\sigma=\nu_{r+1}$ and reordered the indices accordingly.

The cup product for $\mathbb{Z}$ valued $p$-form $\alpha$ and $q$-form $\beta$ is defined by
\begin{align}
    (\alpha \cup \beta)_{x, \mu_1\cdots \mu_{p+q}}= \sum_{\sigma\in \mathfrak{S}_{p+q} } \epsilon_\sigma \frac{1}{p!} \frac{1}{q!} \alpha_{x, \mu_{\sigma(1)}\cdots \mu_{\sigma(p)}  } \beta_{x + \sum_{i=1}^p  \hat{ \mu}_{\sigma(i)}, \mu_{\sigma(p+1)}\cdots \mu_{\sigma(p+q)}  },
\end{align}
where $\mathfrak{S}_{p+q}$ is the symmetric group of order $(p+q)$, and $\epsilon_\sigma$ is the signature of the permutation $\sigma$. Although the definition includes the fractional number $1/(p! q!)$, the product becomes an integer number due to the antisymmetry of $\alpha$ and $\beta$. The cup product satisfies the Leibniz rule,
\begin{align}
d(\alpha \cup \beta)=d\alpha\cup\beta+(-1)^{p}\alpha\cup d\beta. 
\end{align}

A higher cup product $\cup_1$ on a hypercubic lattice is defined in a general way in \cite{Chen:2021ppt}. We translate it into the notation of a lattice gauge theory as\footnote{Our notation is consistent with \cite{Jacobson:2023cmr}, where we flip the sign in Eqs.~(27) and (28) in  \cite{Chen:2021ppt}}
\begin{align}
    (\alpha\cup_1 \beta)_{x, \mu_1 \cdots\mu_{p+q-1} } = (-1)^{q}\sum_{a=1}^{p+q-1} (-1)^{a} P_{\mu_a} ( P_{\mu_a} \iota_{\mu_a} \alpha \cup P_{\mu_a} \iota_{\mu_a} \beta  )_{x, \mu_1\cdots \mathring{\mu}_a \cdots \mu_{p+q-1}} \label{eq: higher cup product}.
\end{align}
Here, $P_{\mu} $ is a parity transformation, which acts as 
\begin{align}
    P_{\mu} x =&P_{\mu}(x_1 , \cdots x_{\mu -1}, x_{\mu} , x_{\mu+1},\cdots ,x_d ) \nonumber \\
    =&(x_1 , \cdots x_{\mu -1},- x_{\mu} , -x_{\mu +1},\cdots ,- x_d ),
\end{align}
and
\begin{align}
    (P_{\mu} \alpha)_{x, \mu_1 \cdots \mu_p}=&  \alpha_{P_\mu x, \mu_1 \cdots \mu_{b-1}(-\mu_b) \cdots  (-\mu_p)} \nonumber \\
   =& (-1)^{ p-b+1 }\alpha_{ P_{\mu} x- \sum_{i=b}^{ p } \hat{\mu}_i  ,\mu_1 \cdots \mu_p   } 
\end{align}
with $\mu_{b-1}<\mu \leq \mu_{b}$. $\iota_\mu $ is an interior product given by
\begin{align}
    (\iota_{\mu} \alpha)_{x, \mu_1 \cdots \mu_p}= \alpha_{x,\mu \mu_1 \cdots \mu_p }.
\end{align}
The higher cup product also satisfies a Leibniz rule
\begin{align}
    d(\alpha \cup_1 \beta  )= d\alpha \cup_1 \beta +(-1)^{p} \alpha \cup_1 d \beta +(-1)^{ p+q+1 }( \alpha \cup \beta -(-1)^{pq } \beta \cup \alpha   ),
\end{align}
with some additional terms.

To prove this, we need the commutativity between $P_\mu$ and $d$,
\begin{align}
    dP_{\mu_a} = P_{\mu_a} d \label{eq: Pd},
\end{align}
and Cartan's magic formula,
\begin{align}
    d \iota_{\mu} +\iota_{\mu} d= \nabla_\mu \label{eq: Cartan Magic},
\end{align}
where $\nabla_\mu f_x = f_{x+ \hat{\mu}} -f_x$ for a $0$-form $f$.
We skip the derivation of them. The exterior derivative of the higher cup product is given by
\begin{align}
    d(\alpha \cup_1 \beta  )_{x, \mu_1 \cdots \mu_{p+q}}= &\sum_{j=1}^{p+q}(-1) ^{j+1} \nabla_{j} ( \alpha \cup_1 \beta  )_{x, \mu_1 \cdots \mathring{\mu_j} \cdots \mu_{p+q}} \nonumber \\
    =&(-1)^{q}\sum_{j=1}^{p+q}(-1) ^{j+1} \left[ \sum_{a=1}^{j-1} (-1)^{a} \nabla_j(P_{\mu_a} ( P_{\mu_a} \iota_{\mu_a} \alpha \cup P_{\mu_a} \iota_{\mu_a} \beta  ))_{x, \mu_1\cdots \mathring{\mu}_a  \cdots \mathring{\mu}_j  \cdots \mu_{p+q}}  \right. \nonumber \\
    &  \left.+ \sum_{a=j+1}^{p+q} (-1)^{a+1} \nabla_j(P_{\mu_a} ( P_{\mu_a} \iota_{\mu_a} \alpha \cup P_{\mu_a} \iota_{\mu_a} \beta  ))_{x, \mu_1\cdots \mathring{\mu}_a  \cdots \mathring{\mu}_j  \cdots \mu_{p+q}}  \right] \nonumber \\
    =&-(-1)^{q}\sum_{a=1}^{p+q}(-1) ^{a} \left[ \sum_{j=1}^{a-1} (-1)^{j+1} \nabla_j(P_{\mu_a} ( P_{\mu_a} \iota_{\mu_a} \alpha \cup P_{\mu_a} \iota_{\mu_a} \beta  ))_{x, \mu_1\cdots \mathring{\mu}_a  \cdots \mathring{\mu}_j  \cdots \mu_{p+q}}  \right. \nonumber \\
    &  \left.+ \sum_{j=a+1}^{p+q} (-1)^{j} \nabla_j(P_{\mu_a} ( P_{\mu_a} \iota_{\mu_a} \alpha \cup P_{\mu_a} \iota_{\mu_a} \beta  ))_{x, \mu_1\cdots \mathring{\mu}_a  \cdots \mathring{\mu}_j  \cdots \mu_{p+q}}  \right] \nonumber \\
    =& -(-1)^{q}\sum_{a=1}^{p+q}(-1) ^{a} d P_{\mu_a} ( P_{\mu_a} \iota_{\mu_a} \alpha \cup P_{\mu_a} \iota_{\mu_a} \beta  )_{x, \mu_1\cdots \mathring{\mu}_a \cdots \mu_{p+q}}.
\end{align}
Using Eqs.~\eqref{eq: Pd} and \eqref{eq: Cartan Magic}, we get
\begin{align}
     d(\alpha \cup_1 \beta  )_{x, \mu_1 \cdots \mu_{p+q}}=&-(-1)^{q}\sum_{a=1}^{p+q}(-1) ^{a}  P_{\mu_a}  ( P_{\mu_a} d\iota_{\mu_a} \alpha \cup P_{\mu_a} \iota_{\mu_a} \beta  )_{x, \mu_1\cdots \mathring{\mu}_a \cdots \mu_{p+q}}\nonumber \\
     &-(-1)^{p-1}(-1)^{q}\sum_{a=1}^{p+q}(-1) ^{a}  P_{\mu_a} ( P_{\mu_a} \iota_{\mu_a} \alpha \cup P_{\mu_a}d \iota_{\mu_a} \beta  )_{x, \mu_1\cdots \mathring{\mu}_a \cdots \mu_{p+q}} \nonumber \\
     =&(-1)^{q}\sum_{a=1}^{p+q}(-1) ^{a}  P_{\mu_a}  ( P_{\mu_a} \iota_{\mu_a}d \alpha \cup P_{\mu_a} \iota_{\mu_a} \beta  )_{x, \mu_1\cdots \mathring{\mu}_a \cdots \mu_{p+q}}\nonumber \\
     &+(-1)^{p-1}(-1)^{q}\sum_{a=1}^{p+q}(-1) ^{a}  P_{\mu_a} ( P_{\mu_a} \iota_{\mu_a} \alpha \cup P_{\mu_a} \iota_{\mu_a}d \beta  )_{x, \mu_1\cdots \mathring{\mu}_a \cdots \mu_{p+q}}  \nonumber \\
     &-(-1)^{q}\sum_{a=1}^{p+q}(-1) ^{a}  P_{\mu_a}  ( P_{\mu_a} \nabla_{\mu_a} \alpha \cup P_{\mu_a} \iota_{\mu_a} \beta  )_{x, \mu_1\cdots \mathring{\mu}_a \cdots \mu_{p+q}}\nonumber \\
     &-(-1)^{p-1}(-1)^{q}\sum_{a=1}^{p+q}(-1) ^{a}  P_{\mu_a} ( P_{\mu_a} \iota_{\mu_a} \alpha \cup P_{\mu_a}\nabla_{\mu_a} \beta  )_{x, \mu_1\cdots \mathring{\mu}_a \cdots \mu_{p+q}}  \nonumber \\
     =&(d\alpha \cup_1 \beta  +(-1)^p  \alpha \cup_1 d \beta )_{x, \mu_1 \cdots \mu_{p+q}}\nonumber \\
      &-(-1)^{q}\sum_{a=1}^{p+q}(-1) ^{a} \left[  P_{\mu_a}  ( P_{\mu_a} \nabla_{\mu_a} \alpha \cup P_{\mu_a} \iota_{\mu_a} \beta \right.  \nonumber \\
     &\left. -(-1)^{p-1}  P_{\mu_a} ( P_{\mu_a} \iota_{\mu_a} \alpha \cup P_{\mu_a}\nabla_{\mu_a} \beta  ) \right] _{x, \mu_1\cdots \mathring{\mu}_a \cdots \mu_{p+q}} .
\end{align}
By a tough calculation, we find
\begin{align}
    \left[  P_{\mu_a}  ( P_{\mu_a} \nabla_{\mu_a} \alpha \cup P_{\mu_a} \iota_{\mu_a} \beta  +(-1)^{p-1}  P_{\mu_a} ( P_{\mu_a} \iota_{\mu_a} \alpha \cup P_{\mu_a}\nabla_{\mu_a} \beta  ) \right] _{x, \mu_1\cdots \mathring{\mu}_a \cdots \mu_{p+q}} \nonumber \\
    =(-1)^a(-1)^{p-1} \qty[ P_{\mu_a}( P_{\mu_a}  \alpha \cup P_{\mu_a}  \beta)-  P_{\mu_{a+1}}( P_{\mu_{a+1}}  \alpha \cup P_{\mu_{a+1}}  \beta)]_{x, \mu_1 \cdots \mu_{p+q}}.
\end{align}
and
\begin{align}
    d(\alpha \cup_1 \beta  )_{x, \mu_1 \cdots \mu_{p+q}}=&(d\alpha \cup_1 \beta  +(-1)^p  \alpha \cup_1 d \beta )_{x, \mu_1 \cdots \mu_{p+q}}\nonumber  \\
      &-(-1)^{p+q+1}\sum_{a=1}^{p+q}\qty[ P_{\mu_a}( P_{\mu_a}  \alpha \cup P_{\mu_a}  \beta)-  P_{\mu_{a+1}}( P_{\mu_{a+1}}  \alpha \cup P_{\mu_{a+1}}  \beta)]_{x, \mu_1 \cdots \mu_{p+q}} \nonumber \\
    =&(d\alpha \cup_1 \beta  +(-1)^p  \alpha \cup_1 d \beta )_{x, \mu_1 \cdots \mu_{p+q}}\nonumber  \\
    &-(-1)^{p+q+1}\qty[ P_{\mu_1}( P_{\mu_1}  \alpha \cup P_{\mu_1}  \beta)-  P_{\mu_{p+q+1}}( P_{\mu_{p+q+1}}  \alpha \cup P_{\mu_{p+q+1}}  \beta)]_{x, \mu_1 \cdots \mu_{p+q}}.
\end{align}
From the definition of the parity operator, the last two terms are given by
\begin{align}
    P_{\mu_1}( P_{\mu_1}  \alpha \cup P_{\mu_1}  \beta)_{x, \mu_1 \cdots \mu_{p+q}}=&( P_{\mu_1}  \alpha \cup P_{\mu_1}  \beta) _{P_1 x- \sum_{i=1}^{p+q} \hat{\mu}_i, \mu_1 \cdots \mu_{p+q}} \nonumber \\
    =&\sum_\sigma \frac{1}{p!q!} \epsilon_\sigma (P_{\mu_1}  \alpha)_{P_1x- \sum_{i=1}^{p+q} \hat{\mu}_i ,\mu_{\sigma(1)}\cdots\mu_{\sigma(p)} }\nonumber \nonumber \\
    &\times 
    (P_{\mu_1}  \beta)_{P_1x- \sum_{i=1}^{p+q} \hat{\mu}_i  +\sum_{i=1}^p  \hat{ \mu}_{\sigma(i)},\mu_{\sigma(p+1)}\cdots\mu_{\sigma(p+q)} }\nonumber  \\
    =&\sum_\sigma \frac{1}{p!q!} \epsilon_\sigma  \alpha_{x+ \sum_{i=1}^{p+q} \hat{\mu}_i -\sum_{i=1}^p  \hat{ \mu}_{\sigma(i)},\mu_{\sigma(1)}\cdots\mu_{\sigma(p)} }
     \beta_{x,\mu_{\sigma(p+1)}\cdots\mu_{\sigma(p+q)} } \nonumber \\
     =&\sum_\sigma \frac{1}{p!q!} \epsilon_\sigma  \alpha_{x+ \sum_{i=p+1}^{p+q} \hat{\mu}_{\sigma(i) },\mu_{\sigma(1)}\cdots\mu_{\sigma(p)} }
     \beta_{x,\mu_{\sigma(p+1)}\cdots\mu_{\sigma(p+q)} }  \nonumber \\
     =&(-1)^{pq} \beta \cup \alpha
\end{align}
and
\begin{align}
    P_{\mu_{p+q+1}}( P_{\mu_{p+q+1}}  \alpha \cup P_{\mu_{p+q+1}}  \beta)_{x, \mu_1 \cdots \mu_{p+q}}=&( P_{\mu_{p+q+1}}  \alpha \cup P_{\mu_{p+q+1}}  \beta)_{P_{\mu_{p+q+1}}x , \mu_1 \cdots \mu_{p+q}}\nonumber \\
    =&\alpha \cup \beta.
\end{align}
Therefore, the Leibniz rule holds.

\section{Poisson Summation Formula}
\label{App: Poisson}

In the Euclidean formulation, $\mathcal{S}$-duality can be seen using the Poisson summation formula. The Poisson summation formula states that the infinite sum over a function is equal to the infinite sum over its Fourier transform
\begin{align}
    \sum_{n=-\infty}^{\infty} f(n)= \sum_{m=-\infty}^{\infty} F(m),
\end{align}
where $n,m\in\mathbb{Z}$, $f(x)$ is a smooth complex function of $x\in\mathbb{R}$ and $F(k)$ is the Fourier transformation of $f(x)$, defined by $F(k)=\int_{-\infty}^{\infty}f(x)e^{i2\pi kx} dx$. The proof is as follows.
\begin{align}
    \sum_{n=-\infty}^{\infty} f(n)&= \sum_{n=-\infty}^{\infty}(\int _{-\infty}^{\infty}F(k)e^{-i2\pi kn}dk) \nonumber\\
    &=\int_{-\infty}^{\infty}F(k)(\sum_{n=-\infty}^{\infty}e^{-i2\pi kn})dk \nonumber\\
    &=\int_{-\infty}^{\infty}F(k)\sum_{m=-\infty}^{\infty}\delta(k-m) dk \nonumber\\
    &=\sum_{m=-\infty}^{\infty}F(m).
\end{align}

Next, we apply the Poisson summation formula to show that Eq.~\eqref{eq: local partition function of Ae Am} is equivalent to Eq.~\eqref{eq: S duality for local}.
To this end, we consider the Fourier transform of the Boltzmann weight appearing in the partition function \eqref{eq: local partition function of Ae Am}, which is given by
\begin{align}
    F(k)&= \int Dx\exp( - \frac{\beta}{2}(dA^e+2\pi x) M (dA^e+2\pi x)  + i\frac{1}{2\pi} \partial A^m( dA^e+2\pi x)   )e^{2\pi i kx} \nonumber \\
    &= \int Dx \exp( - \frac{\beta}{2}(2\pi x) M (2\pi x)  + i\frac{1}{2\pi} \partial A^m( 2\pi x)+i2\pi k(x-\frac{dA^e}{2\pi})   ) \nonumber\\
    &=\int Dx \exp( - \frac{\beta}{2}(2\pi x-\frac{i}{2\pi \beta}(\partial A^{m})M^{-1}) M (2\pi x- \frac{i}{2\pi \beta}M^{-1}(\partial A^{m})) ) \nonumber\\
    &\times \qquad\qquad\exp(-\frac{\beta}{2(2\pi \beta)^{2}} (\partial A^m+2\pi k)M^{-1}(\partial A^m+ 2\pi k)-ikdA^{e}   )) \nonumber\\
    &=\sqrt{\frac{1}{(2\pi \beta)}}^{6N^{4}}\frac{1}{\sqrt{\det(M)}}\exp( -\frac{\beta}{2(2\pi \beta)^{2}} (\partial A^m+2\pi k)M^{-1}(\partial A^m+ 2\pi k)-i\frac{1}{2\pi}(\partial A^m+ 2\pi k)dA^{e}   ),
\end{align}
where
\begin{align}
    \int Dx = \int_{-\infty}^{\infty} \prod_{x} \prod_{\mu<\nu} {dx_{x,\mu\nu}}.
\end{align}
In the last line, we have used $\partial A^{m}dA^{e}=A^{m}ddA^{e}=0$. Substituting this result into the partition function \eqref{eq: local partition function of Ae Am} and applying the Poisson summation formula, we obtain
\begin{align}
    Z[\beta, \theta]&= \int DA^e DA^m \sum_{\{n\}} \exp( -\frac{\beta}{2}(F^e)^2  - i \theta Q[ F^e ] + i\frac{1}{2\pi} \partial A^m F^e   ) \\
    &=\int DA^e DA^m \sum_{\{n\}} \exp( - \frac{\beta}{2}(dA^e+2\pi n) M (dA^e+2\pi n)  + i\frac{1}{2\pi} \partial A^m( dA^e+2\pi n)   ),\\
    &=\sqrt{\frac{1}{(2\pi \beta)}}^{6N^{4}}\frac{1}{\sqrt{\det(M)}}\int DA^e DA^m \sum_{\{m\}}\\
    &\qquad\qquad\times \exp( -\frac{\beta}{2(2\pi \beta)^{2}} (\partial A^m+2\pi m)M^{-1}(\partial A^m+ 2\pi m)-i\frac{1}{2\pi}(\partial A^m+ 2\pi m)dA^{e}   ),\\
    &=\sqrt{\frac{1}{(2\pi \beta)}}^{6N^{4}}\frac{1}{\sqrt{\det(M)}}\int DA^e DA^m \sum_{\{m\}}\\
    &\qquad\qquad\times \exp( -\frac{\beta}{2(2\pi \beta)^{2}} F^{m}M^{-1}F^{m}-i\frac{1}{2\pi}F^{m}dA^{e}   ),
\end{align}
which establishes the dual representation \eqref{eq: S duality for local}.

\section{$\mathrm{SL}(2,\mathbb{Z})$-structure in the Continuum}
\label{App: continuum}

In this appendix, we review the local derivation of the $\mathrm{SL}(2,\mathbb{Z})$-structure of continuum Maxwell theory.
The action of Maxwell theory with $\theta$-term is given as
\begin{align}
S[A]=\int\frac{\beta}{2}F\wedge*F+\frac{i\theta}{8\pi^2}F\wedge F, \label{eq: Maxwell action}
\end{align}
where $F=dA$ is the field strength satisfying Bianchi identity $dF=0$, $\beta$ is the inverse gauge coupling, and $\theta$ is the $\theta$-angle parameter. To see the $\mathrm{SL}(2,\mathbb{Z})$-structure, we define the complex coupling constant $\tau$ as
\begin{align}
\tau = \frac{\theta}{2\pi}+2\pi i\beta.
\end{align}
The $\mathrm{SL}(2,\mathbb{Z})$-structure is generated by $\mathcal{S}$ and $\mathcal{T}$-duality transformation. The $\mathcal{S}$-transformation is the electric-magnetic duality transformation and the $\mathcal{T}$-transformation is the shift of $\theta$ angle by $2\pi$. These transformations can be described as the modular transformations,
\begin{align}
    \mathcal{S}&: \tau \to - \frac{1}{\tau}\\
    \mathcal{T}&: \tau \to \tau+1.
\end{align}

One can see that the $\mathcal{S}$-duality transformation maps $\tau \rightarrow -\frac{1}{\tau}$ by introducing a Lagrange multiplier $\tilde{A}$;
\begin{align}
S[F,\tilde{A}]
&=\int\frac{\beta}{2}F\wedge*F+\frac{i\theta}{8\pi^2}F\wedge F+\frac{i}{2\pi}\tilde{A}\wedge dF,\\
&=\int\frac{\beta}{2}F\wedge*F+\frac{i\theta}{8\pi^2}F\wedge F+\frac{i}{2\pi}\tilde{F}\wedge F,
\label{eq: action with Lagrange multiplier}
\end{align}
Here we impose the Bianchi identity $dF=0$ by the Lagrange multiplier $\tilde{A}$ and treat $F$ as an independent $2$-form dynamical field. In the second line, we perform integration by parts. Taking the variation of this action with respect to $F$, the equation of motion becomes as follows
\begin{align}
\beta*F+\frac{i\theta}{4\pi^2}F+\frac{i}{2\pi}\tilde{F}=0.
\end{align}
Using the equation of motion, $F$ can be expressed in terms of $\tilde{F}$ as
\begin{align}
F=\frac{-\frac{\theta}{2\pi}}{(\frac{\theta}{2\pi})^{2}+(2\pi \beta)^{2}}\tilde{F}+\frac{-2\pi i\beta}{(\frac{\theta}{2\pi})^{2}+(2\pi\beta)^{2}}*\tilde{F}.
\label{eq: equation of motion of F}
\end{align}
Since the action \eqref{eq: action with Lagrange multiplier} is quadratic in $F$, integrating out $F$ yields the same result as substituting the equation of motion into the action. Substituting \eqref{eq: equation of motion of F} into \eqref{eq: action with Lagrange multiplier} gives
\begin{align}
S[\tilde{A}]&=\int\frac{1}{4\pi}\frac{{2\pi\beta}}{(\frac{\theta}{2\pi})^{2}+({2\pi\beta})^{2}}\tilde{F}\wedge*\tilde{F}+\frac{i}{4\pi}\frac{-\frac{\theta }{2\pi}}{(\frac{\theta}{2\pi})^{2}+({2\pi\beta})^{2}}\tilde{F}\wedge\tilde{F},\\
&=\int\frac{\tilde{\beta}}{2}\tilde{F}\wedge*\tilde{F}+\frac{i\tilde{\theta}}{8\pi^2}\tilde{F}\wedge \tilde{F},
\end{align}
where $\tilde{\theta}$ and $\tilde{\beta}$ are defined as
\begin{align}
&\frac{\tilde{\theta}}{2\pi}=\frac{-\frac{\theta }{2\pi}}{(\frac{\theta}{2\pi})^{2}+({2\pi\beta})^{2}},\qquad~2\pi \tilde{\beta}=\frac{2\pi \beta}{(\frac{\theta}{2\pi})^{2}+({2\pi\beta})^{2}} .
\end{align}
The action obtained after integrating out $F$ coincides with the original action \eqref{eq: Maxwell action}, and the transformation of the complex coupling constant under $\mathcal{S}$-duality transformation agrees with $\tau \rightarrow -\frac{1}{\tau}$.

The action of Maxwell theory has the $\mathcal{T}$-duality transformation which is the shift of the $\theta$ angle by $2\pi$. This is because, on a spin manifold, $\frac{1}{8\pi^{2}}\int F\wedge F\in \mathbb{Z}$ and therefore the partition function is invariant under shift $\theta\rightarrow \theta+2\pi$.

\section{Exact Computation of Path Integral}
\label{App: partition function}

In this appendix, we implement the path integral and show that the partition function in Eq.~\eqref{eq: local partition function of Ae Am} can be expressed by theta functions with characteristics. To do this, we use the method based on \cite{Peng:2025nfa}. Specifically speaking, lifting some compact variables into non-compact variables, we carry out the path integral.

\begin{figure}
    \centering
    \includegraphics[width=\linewidth,bb=0 0 860 685]{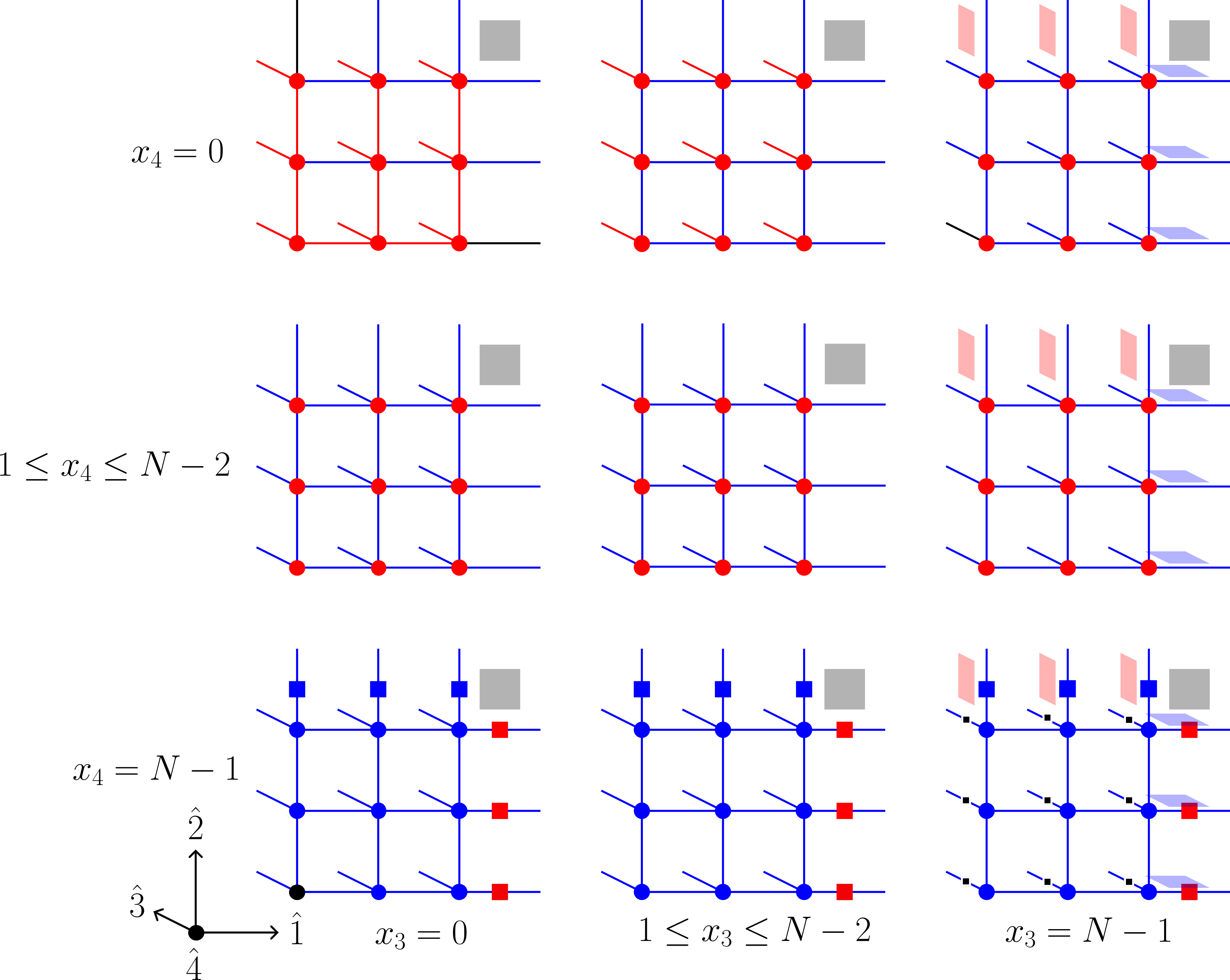}
    \caption{$1$-form gauge field $A^e$ and $2$-form gauge field $n$ after gauge fixing. The nine layers represent the $(x_1, x_2)$ plains with fixed $x_3$ and $x_4$. $A_4^e$ and $n_{\mu4}$ are expressed by filled circles and filled squares on links in the $\mu$ direction. The $2$-form $n$ takes the same value on the plaquettes marked by the same type of filled squares. The $1$-form gauge field $A^e$ on the black links are still compact. }  
    \label{fig: gauge fixing}
\end{figure}

We begin by integrating $A^m$, and use the closedness condition Eq.~\eqref{eq: closedness condition},
\begin{align}
    Z[\beta,\theta]=\int DA^e \sum_{\{n\}} \exp( -\frac{\beta}{2} \sum_x \sum_{\mu<\nu} (F^e)_{x,\mu\nu}^2  - i \theta Q[ F^e ]   ) \prod_x \delta(dn_x)\\
    =\int DA^e \sum_{\{n\}} \exp( -\frac{\beta}{2} \sum_x \sum_{\mu<\nu} (dA^e+ 2\pi n)^2_{x,\mu\nu}  - i \theta \frac{1}{2} \sum_x P_0 n \cup P_0 n   ) \prod_x \delta(dn_x).
\end{align}
This theory poses two gauge symmetries,
\begin{align}
   \mathbb{R}\text{ $0$-form symmetry: } &A^e\to A^e + d\lambda^{(0)},\\
    \mathbb{Z}\text{ $1$-form symmetry: }&A^e\to A^e+ k^{(1)},~n\to n-dk^{(1)}
\end{align}
with $\lambda^{(0)}\in \mathbb{R}$ and $k^{(1)} \in \mathbb{Z}$. According to the argument in \cite{Peng:2025nfa}, $\text{dim}(\text{Im}(d^{(0)}))=N^4-1$ link variables can be eliminated by $0$-form symmetry, and $\text{dim}(\text{Im}(d^{(1)}))=3(N^4-1)$ link variables can be lifted to non-compact variables by $1$-form symmetry by absorbing $3(N^4-1)$ plaquette variables. We show them in Fig.~\ref{fig: gauge fixing}, where link variables on red links are eliminated, and that on blue links are lifted. The filled circle represents $A_{x,4}$. The remaining four link variables are still compact and are depicted as black links. The plaquette variables are almost fixed to zero except for the rectangle symbols. Here, $n_{i4}$ is expressed by rectangles on links. The plaquette variables on the same type of symbol take the same values.

We give a sketch how to remove the $N^4-1$ link variables. Assuming $\lambda^{(0)}_{(1,0,0,0)}=- A^e_{(0,0,0,0),1}$ and the others are zero, then the link variables become
\begin{align}
    A^e_{(0,0,0,0), 1} \to 0 
\end{align}
and 
\begin{align}
    A^e_{ (1,0,0,0),1 } &\to A^e_{ (1,0,0,0),1 }+ A^e_{ (0,0,0,0),1 }, ~\cdots.
\end{align}
Defining new coordinates by the right hand sides, $A^e_{(0,0,0,0), 1}$ disappears from the integrand. We repeat this algorithm and $A^e_{(1,0,0,0), 1},\cdots, A^e_{(N-2,0,0,0), 1}$ are all fixed at zero. However, we cannot fix $A^e_{(N-1,0,0,0),1}$ to zero since it becomes a winding number 
\begin{align}
     A^e_{(N-1,0,0,0),1} \to \sum_{x_1} A^e_{(x_1,0,0,0),1} ,
\end{align}
which is a gauge invariant quantity. We then also apply this for $A^e_{(x_1,x_2,0,0),2},~A^e_{(x_1,x_2,x_3,0),3}$, and $A^e_{(x_1,x_2,x_3,x_4),4}$ and fix the link variables on the red links to zero. Although the Jacobian is unit under this algorithm, the integral range changes from the hypercube to the hyper-parallelogram.

Under the closedness condition, the $6 N^4$ degrees of freedom of $n$ reduce to $ 3(N^4 -1)+ 6 $, where $3(N^4-1)$ is a dimension of $\text{Im}(d^{(1)})$ and six is that of $H^{(2)}$. Then, the integer valued $2$-form $n$ can be decomposed as
\begin{align}
    n= dg+\check{h}.
\end{align}
$g$ represents $\mathbb{Z}$-valued link variables on the blue links, and $\check{h}$ are $\mathbb{Z}$-valued plaquette variables on the rectangle symbols. For example, $\check{h}_{x, 12}$ is determined by
\begin{align}
    \check{h}_{x, 12}=\begin{cases}
        \sum_{ y_1 y_2} n_{(y_1,y_2,x_3,x_4),12}=n^\prime_{12}(0) & (x_1=x_2=N-1) \\
        0 & (\text{Ohterwise}).
    \end{cases}
\end{align}
By absorbing them in $A^e$, $3(N^4-1)$ link variables $A^e$ lift up to non-compact variables $A^\infty$. Then, we finally fix the gauge as presented in Fig.~\ref{fig: gauge fixing}.

Next, we consider the remaining compact gauge field $A^e$ on black links. These gauge field can be removed by the $\mathbb{R}$ $1$-form global symmetry,
\begin{align}
     A^e\to A^e+ \lambda ^{(1)}
\end{align}
with $\lambda^{(1)}\in \mathbb{R}$ and $ d \lambda^{(1)}=0$. Setting 
\begin{align}
    \lambda^{(1)}_{x,\mu}= -A^e_{(N-1)\hat{\mu} ,\mu } ,
\end{align}
we get
\begin{align}
 A^e_{x,\mu}&\to  0\\
    A^\infty_{x,\mu }&\to A^\infty_{x,\mu}- A^e_{(N-1)\hat{\mu},\mu},
\end{align}
where $A^e$ and $A^\infty$ live on the black and blue links. Then, all the compact gauge fields are absorbed into the non-compact gauge fields.

The above argument indicates that the path integral can be expressed as
\begin{align}
    Z[\beta,\theta]=
    \int \qty(\prod_{l: \text{red}} \frac{dA^e_l}{2\pi}) \qty(\prod_{l: \text{black}} \frac{dA^e_l}{2\pi}) \qty(\prod_{l: \text{blue}} \frac{dA^\infty_l}{2\pi})
    \sum_{\{\check{h}\}} \nonumber \\
    \times \exp( - \sum_x \qty[  \frac{\beta}{2}(dA^\infty+ 2\pi \check{h})^2  + i \theta \frac{1}{2}P_0 \check{h} \cup P_0 \check{h}]   ) .
\end{align}
Using the Hodge decomposition,
\begin{align}
    \check{h}= P_d \check{h} + P_0 \check{h}= P_d \check{h} + P_0 n.
\end{align}
we can absorb $ P_d \check{h}$ into $A^\infty$ since $P_d \check{h} \in \text{Im}(d^{(1)})$. Noting $\sum_x d A^\infty P_0n=0$, we get
\begin{align}
    Z[\beta,\theta]=&
    \int \qty(\prod_{l: \text{red}} \frac{dA^e_l}{2\pi}) \qty(\prod_{l: \text{black}} \frac{dA^e_l}{2\pi}) \qty(\prod_{l: \text{blue}} \frac{dA^\infty_l}{2\pi})
    \sum_{\{\check{h}\}} \nonumber \\
    &\times \exp( -\sum_x \qty[ \frac{\beta}{2} (dA^\infty)^2+ \frac{\beta}{2} 4\pi^2 (P_0 n)^2  + i \theta \frac{1}{2}P_0 n \cup P_0 n]   )\\
    =& C \frac{1}{\sqrt{\beta}^{3(N^4-1)}} \prod_{\mu<\nu} \sum_{{n}^\prime_{\mu\nu} }  \exp( - \frac{\beta}{2} 4\pi^2 \sum_{\mu<\nu}(n^\prime_{\mu\nu }(0))^2  - i \frac{1}{2}\theta\sum_{\substack{ \mu< \nu \\ \rho< \sigma }} n^\prime_{\mu\nu }(0) \epsilon_{\mu\nu\rho \sigma}n^\prime_{\rho \sigma }(0)  )\\
    =& C \frac{1}{\sqrt{\beta}^{3(N^4-1)}}\qty(  \sum_{n_1}\sum_{n_2} \exp( - \frac{\beta}{2} 4\pi^2(n_1^2+n_2^2)  - i \theta n_1 n_2 ))^3,
\end{align}
where $C$ is a constant parameter determined by $N$.

\begin{figure}
    \centering
    \includegraphics[width=0.5\linewidth,bb=0 0 228 198]{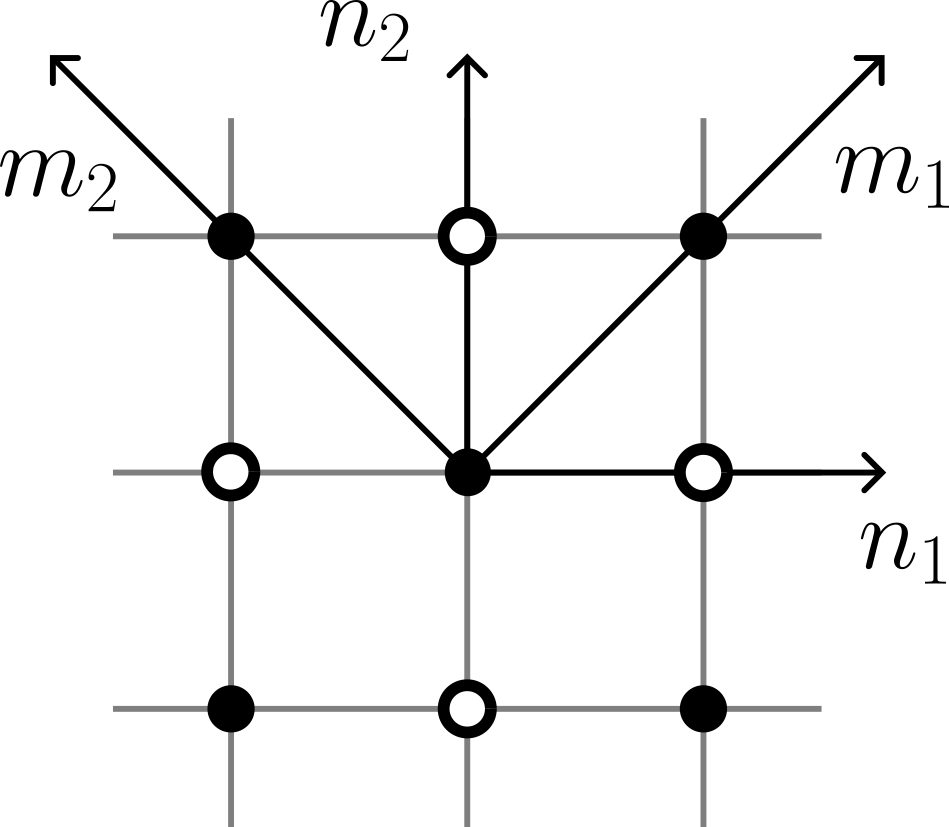}
    \caption{Sublattices $\Lambda_\bullet$ and $\Lambda_\circ$ in two-dimensional lattice $\mathbb{Z}^2$ with the coordinates $(n_1,n_2)$. The filled and open sites belong to $\Lambda_\bullet$ and $\Lambda_\circ$, and are labeled by $(m_1, m_2)$ as in Eqs.~\eqref{eq: bullet site} and \eqref{eq: open site}.  }
    \label{fig: TwoLattice}
\end{figure}

We can rewrite the last equation in terms of theta functions with characteristics defined in Eq.~\eqref{eq: theta function}. To remove the interaction term between $n_1$ and $n_2$, we divide the two dimensional $\mathbb{Z}^2$ lattice into two sublattices,
\begin{align}
    \Lambda_{\bullet} &= \left\{ \mqty( m_1-m_2 \\m_1+m_2 ) \in \mathbb{Z}^2 \;\middle|\; m_1,~m_2 \in \mathbb{Z} \right\}, \label{eq: bullet site}\\
    \Lambda_{\circ}&= \left\{ \mqty( m_1-m_2+1 \\m_1+m_2 ) \in \mathbb{Z}^2 \;\middle|\; m_1,~m_2 \in \mathbb{Z} \right\} \label{eq: open site},
\end{align}
which are depicted in Fig.~\ref{fig: TwoLattice}. The filled and open circles represent sites of $\Lambda_{\bullet}$ and $ \Lambda_{\circ}$, respectively.

The summation on $\Lambda_{\bullet} $ is given by
\begin{align}
&\sum_{(n_1,n_2) \in \Lambda_{\bullet}   } \exp( - \frac{\beta}{2} 4\pi^2(n_1^2+n_2^2)  - i \theta n_1 n_2 ) \\
&=\sum_{(m_1,m_2) \in \mathbb{Z}^2 } \exp( -\frac{4\pi^2\beta}{2} ((m_1+m_2)^2 +(m_1 -m_2)^2 ) -i\theta (m_1+m_2)(m_1-m_2) ) \\
&= \sum_{(m_1,m_2) \in \mathbb{Z}^2 } \exp( -i2\pi m_1^2 \bar{\tau }  + i2\pi m_2^2 \tau   ) \\
&=\abs{ \vartheta\mqty[ 0\\0 ] (0,2\tau) }^2
\end{align}
The calculation on $\Lambda_{\circ} $ is analogous, 
\begin{align}
&\sum_{(n_1,n_2) \in \Lambda_{\circ}   } \exp( - \frac{\beta}{2} 4\pi^2(n_1^2+n_2^2)  - i \theta n_1 n_2 ) 
=\abs{ \vartheta\mqty[ 1/2\\0 ] (0,2\tau) }^2.
\end{align}
Thus, we obtain
\begin{align}
    Z[\beta,\theta]= \frac{C}{\sqrt{\beta}^{3(N^4-1)}} \qty( ~\abs{ \vartheta\mqty[ 0\\0 ] (0,2\tau) }^2 +\abs{ \vartheta\mqty[ 1/2\\0 ] (0,2\tau) }^2~    )^3.
\end{align}

This method also works in the local model presented in \cite{Anosova:2022cjm}. We introduce a new parameter $\gamma$. The partition function is defined by
\begin{align}
    Z[\gamma, \theta]= \int DA^e DA^m \sum_n \exp( -\frac{\beta}{2} \sum_x  F^e_{x,\mu\nu} K_{x,\mu\nu| y,\rho\sigma} F^e_{y,\rho \sigma } -i A^m \cdot dn  ),
\end{align}
where $K$ is a non-ultra-local matrix
\begin{align}
     K= H^{-1/2}  ( 1 + i\gamma \frac{\epsilon + {}^t\epsilon  }{2}  ) 
\end{align}
with
\begin{align}
    H= 1 + \frac{\gamma^2}{4} ( 2+ T_{\hat{s}} + T_{\hat{s}}^{-1} ).
\end{align}
This model is self-dual. By applying the Poisson summation formula, $K$ changes as
\begin{align}
    \beta K\to \frac{1}{4\pi^2 \beta} K^{-1} =  \frac{1}{4\pi^2 \beta} H^{-1/2}  ( 1 - i\gamma \frac{\epsilon + {}^t\epsilon  }{2}). 
\end{align}
That is, the Poisson summation transforms
\begin{align}
    \beta \to \frac{1}{4\pi^2 \beta} ,~\gamma \to \gamma.
\end{align}

$\gamma$ plays the role of $\theta$. Assuming the imaginary part of the quadratic part is the $\theta$-term, we get
\begin{align}
    \theta Q_0 = F^e \cdot  \frac{\beta}{2} P_0 H^{-1/2} \gamma \frac{\epsilon + {}^t\epsilon}{2} P_0 F^e=    \frac{\beta}{2}   \frac{\gamma}{\sqrt{ 1+ \gamma^2}}  F^e \cdot P_0 \frac{\epsilon + {}^t\epsilon}{2} P_0 F^e.
\end{align}
Since $Q_0$ can also be expressed in terms of the harmonic part, $\theta$ is rewritten by $\beta$ and $\gamma$,
\begin{align}
    \theta=  4\pi^2 \beta    \frac{\gamma}{\sqrt{ 1+ \gamma^2}}.
\end{align}
By rescaling $\beta^\prime= \frac{\beta}{ \sqrt{1+\gamma^2}}$\footnote{In the notation of Ref.~\cite{Anosova:2022cjm}, $\beta^\prime= \frac{1}{e^2}$}, we get
\begin{align}
\gamma= \frac{\theta}{4\pi^2 \beta^\prime}. \label{eq: self-dual condtion}
\end{align}

To consider the $\mathrm{SL}(2,\mathbb{Z})$-duality, it is convenient to define 
\begin{align}
    Z[\gamma, \beta^\prime,\theta]=& \int DA^e DA^m \sum_n \exp( -\frac{\beta^\prime }{2} \sum_{x,y} \sum_{\mu<\nu}  F^e_{x,\mu\nu}  \frac{\sqrt{1+\gamma^2}}{\sqrt{H}_{x,y}} F^e_{y,\mu\nu } - i\theta Q_0[F^e]-  i A^m \cdot dn  ) ,
\end{align}
where we regard $\gamma, \beta, \theta$ as independent parameters. When $\gamma$ satisfies the condition~\eqref{eq: self-dual condtion}, this partition function is self-dual from the above argument,
\begin{align}
    Z\qty[\gamma= \frac{\theta}{4\pi^2 \beta^\prime} , \beta^\prime,\theta]=Z\qty[\gamma= \frac{\theta}{4\pi^2 \beta^\prime} , \tilde{\beta}^\prime,\tilde{\theta}] =Z\qty[\gamma= \frac{ \tilde{\theta}}{4\pi^2 \tilde{\beta}^\prime} , \tilde{\beta}^\prime,\tilde{\theta}],
\end{align}
where $\tilde{\beta}^\prime$ and $\tilde{\theta}$ are defined by replacing $\beta$ by $\tilde{\beta}^\prime$ in Eqs.~\eqref{eq: tilde beta} and \eqref{eq: tilde theta}. 

Since $Q_0$ is an integer, the periodicity of $\theta$ is $2\pi$,
\begin{align}
    Z[\gamma, \beta^\prime,\theta+ 2\pi]=Z[\gamma, \beta^\prime,\theta].
\end{align}
However, shifting $\theta$ violates the self-dual condition \eqref{eq: self-dual condtion},
\begin{align}
    \gamma= \frac{\theta }{4\pi^2 \beta^\prime } \neq \frac{\theta+ 2\pi }{4\pi^2 \beta^\prime }.
\end{align}
That is, it is non-trivial to reconcile them.  

We calculate the path integral explicitly. By applying the gauge fixing procedure, we obtain
\begin{align}
    Z[\gamma, \beta^\prime,\theta]
    =&    \int \qty(\prod_{l: \text{red}} \frac{dA^e_l}{2\pi}) \qty(\prod_{l: \text{black}} \frac{dA^e_l}{2\pi}) \qty(\prod_{l: \text{blue}} \frac{dA^\infty_l}{2\pi})
    \sum_{\{\check{h}\}} \nonumber \\
    &\times \exp( -\sum_x \qty[ \frac{\beta^\prime}{2} dA^\infty \frac{\sqrt{1+\gamma^2}}{\sqrt{H}} dA^{\infty}  + \frac{\beta}{2} 4\pi^2 (P_0 n)^2  + i \theta \frac{1}{2}P_0 n \cup P_0 n]   )\\
    =& C \frac{ \sqrt{\det^\prime  \frac{\sqrt{H}}{ \sqrt{1+\gamma^2}}} 
    }{\sqrt{\beta^\prime}^{3(N^4-1)}} \prod_{\mu<\nu} \sum_{{n}^\prime_{\mu\nu} }  \exp( - \frac{\beta^\prime}{2} 4\pi^2 \sum_{\mu<\nu}(n^\prime_{\mu\nu }(0))^2  - i \frac{1}{2}\theta\sum_{\substack{ \mu< \nu \\ \rho< \sigma }} n^\prime_{\mu\nu }(0) \epsilon_{\mu\nu\rho \sigma}n^\prime_{\rho \sigma }(0)  ).
\end{align}
Here, $\det^\prime$ means that a matrix is projected onto the blue links. Naively, 
\begin{align}
   \mathrm{det}^\prime\qty( \frac{\sqrt{H}}{\sqrt{1+ \gamma^2}}) \propto \sqrt{ \prod_{p} \qty( \frac{ 1+ \gamma^2 \frac{1+ \cos(p_s)  }{2}}{1+ \gamma^2}  )^3    }
\end{align}
and this is not invariant under shifting $\theta$. To restore the periodicity of $\theta$, we need Pauli-Villars fields. Even with this prescription, the $\mathrm{SL}(2,\mathbb{Z})$-duality in the presence of loop operators requires further discussion.

\bibliographystyle{JHEP}
\bibliography{ref}


\end{document}